\documentclass[twocolumn,tighten,times]{aastex63}
\usepackage{amsmath}

\date{\today}
\received{May 29, 2019}
\accepted{April 24, 2020}
\submitjournal{the Astrophysical Journal}

\def\gtsim {>\kern-1.2em\lower1.1ex\hbox{$\sim$}~}   % Greater than sim
\def\ltsim {<\kern-1.2em\lower1.1ex\hbox{$\sim$}~}   % Less than sim

\begin{document}

\title{New Type Ia Supernova Yields and the Manganese and Nickel Problems in the Milky Way and Dwarf Spheroidal Galaxies}

%\correspondingauthor{Chiaki Kobayashi}
%\email{c.kobayashi@herts.ac.uk}
%, shingchi.leung@ipmu.jp, nomoto@astron.s.u-tokyo.ac.jp

\author{Chiaki Kobayashi}
\affiliation{School of Physics, Astronomy and Mathematics, Centre for Astrophysics Research, University of Hertfordshire, College Lane, Hatfield AL10 9AB, UK}
\affiliation{Kavli Institute for the Physics and 
Mathematics of the Universe (WPI), The University 
of Tokyo Institutes for Advanced Study, The 
University of Tokyo, Kashiwa, Chiba 277-8583, Japan}
\author{Shing-Chi Leung}
\affiliation{Kavli Institute for the Physics and 
Mathematics of the Universe (WPI), The University 
of Tokyo Institutes for Advanced Study, The 
University of Tokyo, Kashiwa, Chiba 277-8583, Japan}
\affiliation{TAPIR, Walter Burke Institute for Theoretical Physics, 
Mailcode 350-17, Caltech, Pasadena, CA 91125, USA}
\author{Ken'ichi Nomoto}
\affiliation{Kavli Institute for the Physics and 
Mathematics of the Universe (WPI), The University 
of Tokyo Institutes for Advanced Study, The 
University of Tokyo, Kashiwa, Chiba 277-8583, Japan}

%\correspondingauthor{Chiaki Kobayashi}
%\email{c.kobayashi@herts.ac.uk}

\begin{abstract}
In our quest to identify the progenitors of Type Ia supernovae (SNe Ia), we first update the nucleosynthesis yields both for near-Chandrasekhar (Ch) and sub-Ch mass white dwarfs (WDs), for a wide range of metallicity, with our two-dimensional hydrodynamical code and the latest nuclear reaction rates.
We then include the yields in our galactic chemical evolution code to predict the evolution of elemental abundances in the solar neighborhood and dwarf spheroidal (dSph) galaxies: Fornax, Sculptor, Sextans, and Carina.
In the observations of the solar neighborhood stars, Mn shows an opposite trend to $\alpha$ elements, showing an increase toward higher metallicities, which is very well reproduced by deflagration-detonation transition of Ch-mass WDs, but never by double detonations of sub-Ch-mass WDs alone.
The problem of Ch-mass SNe Ia was the Ni over-production at high metallicities.
However, we found that Ni yields of Ch-mass SNe Ia are much lower with the solar-scaled initial composition than in previous works, which keeps the predicted Ni abundance within the observational scatter.
From the evolutionary trends of elemental abundances in the solar neighborhood, we conclude that the contribution of sub-Ch-mass SNe Ia in chemical enrichment is up to 25\%.
In dSph galaxies, however, larger enrichment from sub-Ch-mass SNe Ia than in the solar neighborhood may be required, which causes a decrease in [(Mg, Cr, Mn, Ni)/Fe] at lower metallicities.
The observed high [Mn/Fe] ratios in Sculptor and Carina may also require additional enrichment from pure deflagrations, possibly as Type Iax supernovae.
Future observations of dSph stars will provide more stringent constraints on the progenitor systems and explosion mechanism of SNe Ia.
\end{abstract}



\section{Introduction}
\label{sec:intro}
Although Type Ia Supernovae (SNe Ia) have been used as a standard candle to measure the expansion of the Universe, there is a small but significant variation in their luminosities. Brighter SNe Ia show a slower decay, which shows a correlation between the peak luminosity and light curve width \citep{phi93}. The luminosity variation is empirically corrected in the supernova cosmology \citep{Perlmutter1999,Riess1998}, although the physical origin of the relation is uncertain.
The dependence of this variation on the host galaxies has first been reported by \citet{ham96}, where the mean peak brightness is dimmer in elliptical galaxies than in spiral galaxies.
\citet{ume99} provided the first theoretical explanation for this dependence, assuming that a smaller C/O ratio leads to a dimmer SN Ia (see also visualization in Fig. 7 of \citealt{nom00}).
Similar dependencies of the luminosity variation on various properties of host galaxies are found in more recent observations \citep[e.g.,][]{chi13}, but the origin of the variation has not been confirmed yet.

The progenitor of SNe Ia is still a matter of big debate (see \citealt {hil00,mao14,sok19} for a review). It is a combined problem of the progenitor systems and the explosion mechanism. In recent works, common progenitors are (1) deflagration or delayed detonation (DDT) of a near-Chandrasekhar (Ch)-mass carbon-oxygen (C+O) white dwarf (WD) in a single degenerate system \citep[][]{whe73,Nomoto1982a}, (2) sub-Ch-mass explosion in a double degenerate system \citep[e.g.,][]{ibe84,web84,pak12}, (3) double detonations of sub-Ch-mass WDs in a single or double degenerate system \citep[e,g,][]{Nomoto1982b,ibe91,rui14}, (4) weak deflagration of a near-Ch or super-Ch mass WD with a low mass WD remnant in a single degenerate system, which possibly correspond to a Type Iax supernova \citep[SN Iax,][]{fol13,men14,fin14,mcc14}, and (5) delayed explosion of a rotating super-Ch-mass C+O WD \citep{ben15}, which could be formed from merging of a C+O WD with the core of massive asymptotic giant branch (AGB) star during common envelope evolution \citep{sok15}.

For the nucleosynthesis yields of SNe Ia, the so-called W7 model \citep{Nomoto1984,thi86,Nomoto1997,iwa99} has been the most favoured 1D model for reproducing the observed spectra of SNe Ia \citep{hof96,nug97}.
In recent works, 3D simulations of a delayed detonation in a Ch-mass WD and of a violent merger of two WDs \citep{Roepke2012}, and 2D simulations of a double detonation in a sub-Ch-mass WD \citep{kro10} can also give a reasonable match with observations.
The advantage of the W7 model is that it also reproduces the Galactic chemical evolution (GCE) in the solar neighborhood, namely, the observed increase of Mn/Fe with metallicity as well as the decrease of $\alpha$ elements (O, Mg, Si, S, and Ca) \citep{kob06}; with the updated nucleosynthesis yields of core-collapse supernovae (with a mix of normal supernovae with $10^{51}$ erg and hypernovae with $\ge 10^{52}$ erg at $\ge 20M_\odot$ stars), [Mn/Fe] is about $-0.5$ at [Fe/H] $\ltsim -1$, and increases toward higher metallicities because of the delayed enrichment of SNe Ia. 
However, there is a remaining problem in GCE with the W7 yields; the Ni/Fe ratio is higher than observed at [Fe/H] $\gtsim -1$, which could be solved with DDT models \citep[e.g.,][]{iwa99}.
An updated GCE model with the DDT yields from \citet{Seitenzahl2013} was shown in \citet{sne16}, which indeed gives Ni/Fe ratios closer to the observational data.

In contrast to these Ch-mass models, sub-Ch-mass models, which have been re-considered for SNe Ia with a number of other observational results such as supernova rates \citep[e.g.,][]{mao14} and the lack of donors in supernovae remnants \citep{ker09}, do not match the GCE in the solar neighborhood.
The Mn production from sub-Ch-mass models is too small to explain the observations in the solar neighborhood \citep{Seitenzahl2013}.
SNe Iax could compensate this with their large Mn production, but their rate seems to be too low for the solar neighborhood \citep[][hereafter K19]{kob15,kob19}.

Recently, dwarf spheroidal galaxies (dSphs) have been used as another site for constraining nucleosynthesis yields because of their low metallicities.
Using our GCE model, \citet{kob15} showed that a mix of sub-Ch-mass SNe Ia and SNe Iax may be able to explain the scatter in the observed abundance ratios, which was confirmed by a stochastic chemical evolution model in \citet{ces17}.
Recently, \citet{kir19} used a large sample of observational data and concluded that sub-Ch-mass SNe Ia are the main enrichment source in dSphs.

In this paper, we test SN Ia progenitor models using updated SN Ia yields sets both for Ch and sub-Ch mass explosions. The yields are calculated with our new 2D explosion and nucleosynthesis code \citep{Leung2018} for a wide range of metallicity (\S 2). 
Using a GCE model \citep{kob00}, we show the evolution of elemental abundances ratios of iron-peak elements in the solar neighborhood (\S 3) and dSphs (\S 4), and put a constraint on the explosion models of SNe Ia comparing with observed stellar abundances.
\S5 gives our conclusions.

\begin{table*}
\begin{center}
\caption{
Model setup for the benchmark models with the initial metallicity $Z = 0.02$:
``Mechanism'' is the explosion mechanism used in our simulations
including the DDT and double detonation (DD) models.
Central densities of  $\rho_{\rm c}$
are in units of $10^{8}$ g cm$^{-3}$.  
The total mass of WDs, $M_{\rm WD}$, and helium envelope mass, $M_{{\rm He}}$, are in units of solar mass. 
$R$ is the initial stellar radius in kilometers. 
$E_{\rm nuc}$ and $E_{\rm tot}$ are the energy released by nuclear reactions
and final total energy, respectively, both in units of $10^{50}$ erg. 
$t_{{\rm trans}}$ is 
the first detonation transition time in units of second.
$M({\rm ^{56}Ni})$, $M({\rm Mn})$, and $M({\rm Mn})$ 
are the masses of $^{56}$Ni,
stable Mn, and $^{58}$Ni at the end of
simulations, after short-live radioactive isotopes 
have decayed.
}
\label{table:std}
\begin{tabular}{|c|c|c|c|c|c|c|c|c|c|c|c|c|c|}
\hline

Model & mechanism  & $\rho_{c({\rm NM})}$ & $M_{\rm WD}$ & $M_{\rm He}$ &
$R$ & $E_{{\rm nuc}}$ & $E_{{\rm tot}}$ & 
$t_{{\rm trans}}$ & $M(^{56}{\rm Ni})$ & $M({\rm Mn})$ & $M(^{58}$Ni) \\ \hline
Near-Ch-mass model & DDT & 30 & 1.38 & 0 & 1900 
& 17.7 & 12.7 & 0.78 & 0.63 & $8.46 \times 10^{-3}$ & $4.42 \times 10^{-2}$ \\ \hline
Sub-Ch-mass model & DD & 0.32 & 1.0 & 0.05 & 6200 & 10.2 & 8.7 & 0.98 & 0.63 & $5.68 \times 10^{-4}$ & $1.34 \times 10^{-3}$ \\ \hline

\end{tabular}
\end{center}
\end{table*}

\section{Nucleosynthesis yields}

Here we briefly describe
the methods for producing the representative
SN Ia models using both Ch and
sub-Ch mass C+O WD.
Detailed model descriptions and parameter studies can be found in \citet[][hereafter LN18]{Leung2018} and \citet[][hereafter LN19]{Leung2019subChand}, respectively.

\subsection{Methods}

We use our own two-dimensional hydrodynamics
code, primarily developed to model SNe Ia 
\citep{Leung2015a}. The code has been applied
to various types of SN explosions, including sub-luminous 
SNe Ia \citep{Leung2015b}, near-Ch-mass 
SNe Ia (\citealt{Nomoto2017a}; LN18)
sub-Ch-mass SNe Ia (LN19)
and electron-capture SNe \citep{Nomoto2017b, Leung2019PASA, Leung2019ECSN}.
The code includes the necessary physics such as
the flame-capturing scheme by the level-set method \citep{Reinecke1999a}
with reinitilization \citep{Sussman1994},
sub-grid turbulence \citep{Clement1993,Niemeyer1995a,Schmidt2006b},
and the three-step simplified nuclear reaction scheme \citep{Calder2007}.
In contrast to \cite{Calder2007}, we choose to record the 
chemical composition in the hydrodynamical simulations
explicitly;
our hydrodynamical code includes
a simplified 7-isotope network
of $^{4}$He, $^{12}$C, $^{16}$O, $^{20}$Ne, 
$^{24}$Mg, $^{28}$Si, and $^{56}$Ni \citep[Eq.8 of LN18, see also][]{Timmes2000a}
with their three-step scheme.

For post-processing nucleosynthesis, we use a larger 495-isotope network
for nuclear reactions, containing isotopes from $^{1}$H to $^{91}$Tc.
We use the tracer particle
scheme \citep{Travaglio2004}, which records the thermodynamic
trajectory $\rho-T$ as a function of time. We also use the 
$torch$ nuclear reaction network \citep{Timmes1999a} to compute the exact nucleosynthesis
yields.
Nucleosynthesis yield tables are obtained after short-life radioactive isotopes
have decayed\footnote{The decay time was $10^3$ years in LN18 and LN19, but is $10^6$ years in this paper, which results in a significant difference in Co yields.}.
Note that $^{26}$Al and $^{60}$Fe yields are added to those of $^{26}$Mg and $^{60}$Ni, respectively, in GCE calculation.

\subsection{Near-Chandrasekhar-Mass White Dwarf}

For near-Ch-mass models, we first construct
an isothermal hydrostatic equilibrium C+O WD.
In this paper,
we assume the central density $\rho_{\rm c} = 3 \times 10^9$ g cm$^{-3}$
with uniform temperature $10^8$ K \citep{Nomoto1982a}.
The composition is assumed to be uniform as 
$X(^{12}$C) = $X(^{16}$O) = $(1-Z)/2$
for the metallicities of $Z= 0, 0.002, 0.01, 0.02, 0.04, 0.06,$ and $0.10$.
The $Z$ component is scaled to the solar abundances \citep{Lodders2010} in this paper, which gives a significant difference in the nucleosynthesis yields.
With $Z=0.02$,
the benchmark model is selected by 
requiring three conditions: 1) it has
a yield of $^{56}$Ni $\sim$ 0.6 $M_{\odot}$ as found in typical SNe Ia \citep{Li2011, Piro2014};
2) it has a comparable Mn production at the solar metallicity; 3) it does not severely
over-produce stable Ni.

In Table \ref{table:std} we tabulate
the fundamental stellar parameters and
the resultant explosion energies of our benchmark models for 
Ch and sub-Ch mass SNe Ia. It can be seen 
that the nucleosynthesis yields of the near-Ch-mass model satisfy
these three criteria of the benchmark models.

In LN18 we have computed
45 models of SNe Ia using near-Ch-mass 
C+O WDs as the progenitors.
In view of the diversity of observed
SNe Ia, an extended 
parameter space, including a central densities
of $5 \times 10^8$ to $5 \times 10^9$ g cm$^{-3}$
(corresponding to initial masses of 1.30 - 1.38
$M_\odot$), metallicities from $X$($^{22}$Ne) $=$ 0 to 5 $Z_\odot$\footnote{The solar metallicity was 0.02 and the other elements were not included in the initial composition in LN18 and LN19, which gives a significant difference in $^{58}$Ni yields.},
C/O mass ratios from 0.3 to 1, and different 
ignition kernels from the centered flame to the
off-centered flame have been surveyed. We have then shown that 
the central density and metallicity are important 
parameters that strongly affect 
nucleosynthesis yields;
higher central
density allows larger production in
neutron-rich isotopes such as $^{50}$Ti, $^{54}$Cr, 
$^{58}$Fe, and $^{64}$Ni, while higher
metallicity mostly enhances isotopes related to
the direct product of $^{22}$Ne, such as
$^{50}$V, $^{50}$Cr, $^{54}$Fe, and $^{58}$Ni.

For the explosion mechanism, in this paper
the turbulent deflagration
model with deflagration-detonation transition (DDT) \citep[see e.g.][]{Khokhlov1991a, Golombek2005, Roepke2007d, Seitenzahl2013}
is adopted
for the following two reasons. 
First, the multi-dimensional pure turbulent deflagration (PTD) model \citep{Reinecke1999b, Reinecke2002a, Reinecke2002b, Roepke2007a, Ma2013, fin14} is very likely to leave a remnant and its
explosion is weak. The low ejecta mass may not be important
for chemical enrichment compared to other explosion 
models.
Second, the gravitationally confined detonation (GCD) model \citep{Plewa2004, Jordan2008, Meakin2009, Jordan2012, Seitenzahl2016} tends to produce very strong explosions
with a small amount of neutron-rich isotopes, including Mn.
As discussed in \cite{Seitenzahl2013}, there is not yet another
major site for the production of Mn. Therefore, we focus on the DDT
model, which is more robust in producing iron-peak elements,
although the PTD was also investigated in LN18.

In the core of near-Ch-mass C+O WDs, we introduce an initial
carbon deflagration. The flame structure is 
a ``three-finger'' structure as in \cite{Reinecke1999b}.
Other flame structures were also investigated in 
LN18 and we showed that the overall abundance
pattern is less sensitive to the initial flame structure.

The deflagration starts at the center of the WD
and makes the star expand slowly,
so that the core is always the place of highest
central density and temperature. 
At $t \sim 1$ second after the deflagration started,
the DDT occurs. The detonation provides
a strong shock for compressing the surrounding
material.
This causes a sharp rise in the global maximum density and temperature ($\rho_{\rm max}$ and $T_{\rm max}$, respectively),
which leads to a ``wiggling'' rise in the central density and temperature from 1 - 2 seconds.
Beyond $t \sim 10$ seconds, 
the star enters homologous expansion and 
observable exothermic nuclear reactions take place
(see Figs. 2, 3, 25, and 26 of LN18 for the density, temperature, energy, and luminosity evolution).

\begin{figure}
\centering
\includegraphics*[width=8cm,height=5.7cm]{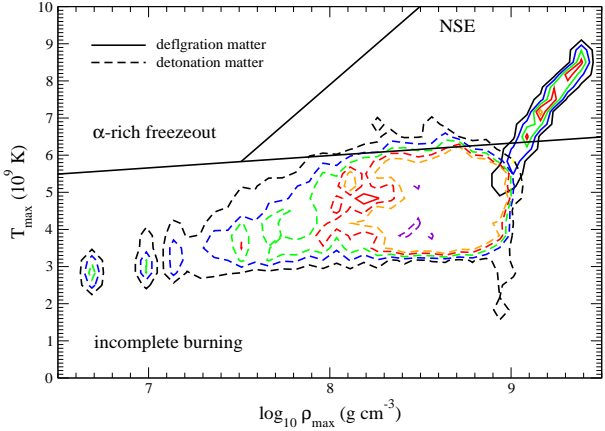}
\caption{$T_{{\rm max}}$ against $\rho_{{\rm max}}$ 
for the near-Ch-mass benchmark model according to 
the thermodynamic trajectories. The tracer particles
being burnt by deflagration (solid lines) or detonation (dashed lines) are separated. Contours stand for 
100 (black), 300 (blue), 500 (green), 700 
(red), and 900 (orange) tracer particles, respectively.
The straight lines roughly indicate the nuclear reactions in this diagram; nuclear statistical equilibrium (NSE), $\alpha$-rich freezeout, and incomplete Si-burning \citep{woo73}. % Eq.67
}
\label{fig:traj_Chand_std}
\end{figure}

In Figure \ref{fig:traj_Chand_std} we show the distribution of
$T_{{\rm max}}$ against $\rho_{{\rm max}}$ for
the near-Ch-mass benchmark model according
to the thermodynamic trajectories of the tracer particles.
There are two populations of tracer particles.
For $\rho_{{\rm max}} \geqslant 10^9$ g cm$^{-3}$, 
there is a tight relation of $T_{{\rm max}}$ increasing with $\rho_{{\rm max}}$.
This corresponds to the 
particles being incinerated by the deflagration wave.
Due to the sub-sonic nature, no shock wave is created
during its propagation. The particles are burnt 
according to their local density. On the other hand,
for particles with $\rho_{{\rm max}} < 10^9$ g cm$^{-3}$,
$T_{{\rm max}}$ spans a wider range. This corresponds to the 
particles being incinerated by the detonation wave. 
Because there is more than one C-detonation 
triggered during the explosion, the collision of 
shock waves provide an observable shock heating, which 
creates the $T_{{\rm max}}$ spectra as seen in the figure. 

\begin{figure}
\centering
\includegraphics*[width=8cm,height=5.7cm]{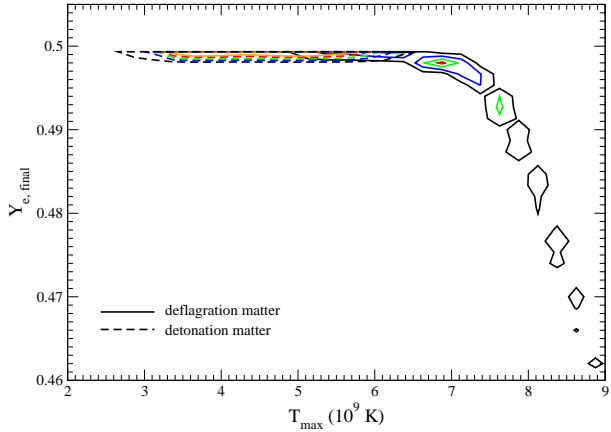}
\caption{$Y_{e,{\rm~min}}$ against $T_{{\rm max}}$ 
for the near-Ch-mass benchmark model according to 
the thermodynamic trajectories. The tracer
particles burnt by deflagration (solid lines) and detonation (dashed lines) are separated. 
The contours are the same as in
Figure \ref{fig:ye_Chand_std}.}
\label{fig:ye_Chand_std}
\end{figure}

In Figure \ref{fig:ye_Chand_std} we show the distribution of
the electron fraction, $Y_e$, against $T_{{\rm max}}$ for the 
tracer particles. It can be seen again, that there are two 
populations of particles. At
$T_{{\rm max}} > 7 \times 10^9$ K, $Y_e$
drops towards higher $T_{{\rm max}}$
from the initial 0.5 to $\sim 0.46$.
This corresponds to
the particles incinerated by the deflagration wave
at high densities, where electron capture can 
efficiently take place. The other population corresponds
to the particles burnt by the detonation wave
or by the deflagration wave with a low density. 
Electron capture occurs at a much slower rate,
so that $Y_e$ stays between $0.5$ and $0.499$ at $T_{{\rm max}} < 7 \times 10^9$ K. % 2 percent of 22Ne

\begin{figure}
\centering
\includegraphics*[width=8cm,height=5.7cm]{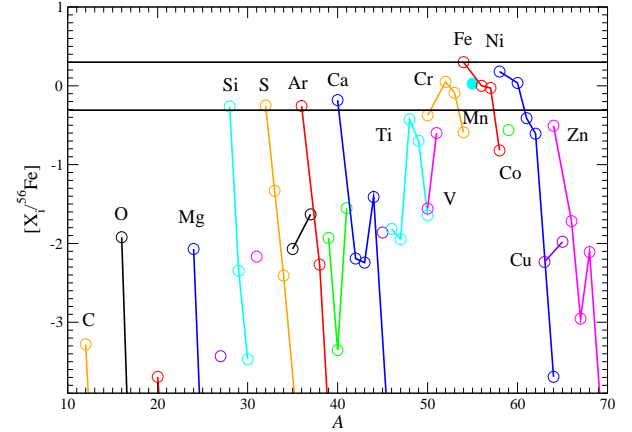}
\caption{$[X_i/^{56}{\rm Fe}]$ of stable isotopes 
in the near-Ch-mass benchmark model 
after short-lived radioactive isotopes have 
decayed. The ratios are scaled to the solar ratios. 
The horizontal lines at
$\pm 0.3$ correspond to 0.5 or 2.0 times the solar values.}
\label{fig:final_Chand_std}
\end{figure}

The nucleosynthesis yields of the benchmark model
are shown 
in Figure \ref{fig:final_Chand_std}, where
mass ratios scaled to the solar ratios, [$X_i/^{56}$Fe], 
are plotted against the mass number.
The two horizontal lines correspond to 
twice- and half-solar ratios. 
Due to the fast detonation wave, very small amounts of C, O,
and Ne are left in the WD. The detonation wave 
mostly burns matter at a low density and produces 
intermediate mass elements from Si to Ca close to
the solar ratios. 
One can also see a healthy production of iron peaked 
elements from Cr to Ni, except mild over-production of $^{54}$Fe and
$^{58}$Ni. Heavier iron-peak elements
such as Co and Zn are under-produced.

\begin{figure}
\centering
\includegraphics*[width=8cm,height=5.7cm]{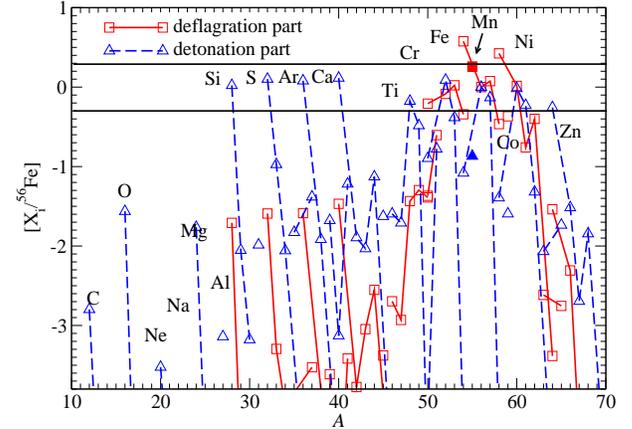}
\caption{The same as Figure \ref{fig:final_Chand_std} but for the particles ignited by deflagration (solid lines) and detonation (dashed lines), respectively.}
\label{fig:final_Chand_defndet_std}
\end{figure}

In Figure \ref{fig:final_Chand_defndet_std} we plot 
the same abundances split into the deflagration 
and detonation components. 
As noted previously, the detonation (dashed line)
mainly burns the low density matter, and small amounts of 
C, O and Ne are left. It mostly produces
the intermediate mass elements, in particular
$^{28}$Si, $^{32}$S, $^{36}$Ar and $^{40}$Ca,
with the values even higher than the solar ratios. The detonation
wave also produces some iron-peak elements
very close to the solar ratios. On the other hand,
the deflagration wave (solid line) burns mainly
high density matter and no fuel is left. It
also produces very little intermediate mass elements. 
However, electron capture
occurs mainly in matter burnt by the deflagration,
where the iron-peak elements, including neutron-rich 
ones such as $^{54}$Fe, $^{55}$Mn, and $^{58}$Ni, are largely enhanced.

\subsection{Sub-Chandrasekhar-Mass White Dwarf}

For the sub-Ch-mass models, we construct a
two-layer WD with carbon-oxygen in the core
and pure helium in the envelope. The helium layer has to be thin (e.g., Fig. \ref{fig:final_sChand_std}),
and in this paper we adopt $M({\rm He})=0.05 ~M_{\odot}$.
Note that this value is smaller than assumed in the binary population synthesis model by \citet[][$0.1M_\odot$]{rui11} and is consistent with other previous works on explosions \citep{bil07,she09,fin10,kro10,woo11}.
The total WD masses including the He layer are 0.9, 0.95, 1.0, $1.1$, and $1.2M_\odot$.
The assumption of the composition is the same as near-Ch-mass models in \S 2.2 but with the metallicities of $Z= 0$, 0.001, 0.002, 0.004, 0.01, $0.02$, and $0.04$.
For $Z=0.02$,
the benchmark model is selected to
produce a normal SN Ia of $^{56}$Ni mass 
$\sim 0.6 ~M_{\odot}$. It is known that
sub-Ch-mass models cannot produce
sufficient Mn for explaining the solar abundance
\citep{Seitenzahl2013}. Thus, we do not 
impose any constraint on the Mn production. 
Again, we also require stable Ni not to be
over-produced.

In LN19 we have computed a series
of 40 models of SNe Ia using sub-Ch
mass C+O WDs as the progenitors. 
A wide range of models with
a progenitor mass from 0.9 - 1.2 $M_\odot$ has been
computed for metallicities from 0 to 5 $Z_\odot$\footnote{The initial composition and the decay time are updated in this paper, similar to near-Ch-mass models.},
C/O mass ratios from 0.3 to 1.0, and He envelope masses from 0.05 to 0.2 $M_\odot$.
The initial mass
and metallicity strongly affects nucleosynthesis yields.
Unlike the near-Ch-mass models where
the central density determines the occurrence of electron capture,
the initial mass determines the $^{56}$Ni production,
and the abundance pattern mainly depends on the scaling with
$^{56}$Fe. 
Therefore, compared to the near-Ch-mass models, 
there is a smaller variety of abundance patterns for sub-Ch-mass models
because of its pure detonation
nature, where most matter does not have a sufficiently high
density for rapid electron capture before it 
cools down by expansion.

For the explosion mechanism, in this paper
the double detonation
model is used, where the carbon detonation is triggered by 
helium detonation.
In LN19, multiple types
of detonation-triggers were investigated; one bubble (a spherical shell), 
multiple bubbles, and a belt-shaped helium detonation at the beginning of the simulations.
Although this affects the minimum helium mass required for detonation, for the models with $M({\rm Fe})\sim0.6M_\odot$ the abundance patterns of iron-peak elements are not so different, and thus we only use the spherical one (``S''-type in LN19) in this paper.

The simulation starts from a He detonation in a 100km spherical shell just outside of the C+O core.
Since it is super-sonic,
both central density and temperature of
the WD remain unchanged for the first 1 second,
although the global maximum temperature ($T_{{\rm max}}$) gradually decreases.
Once the shock wave reaches the centre, the central C-detonation is triggered,
and the central temperature and density rapidly increase.
After that the expansion allows the 
matter to cool down rapidly.
Both central and global maximum densities
drop together, showing that the core has relaxed
and starts its expansion
(see Figs. 5, 6, and 7 of LN19 for the temperature, energy, and luminosity evolution).

\begin{figure}
\centering
\includegraphics*[width=8cm,height=5.7cm]{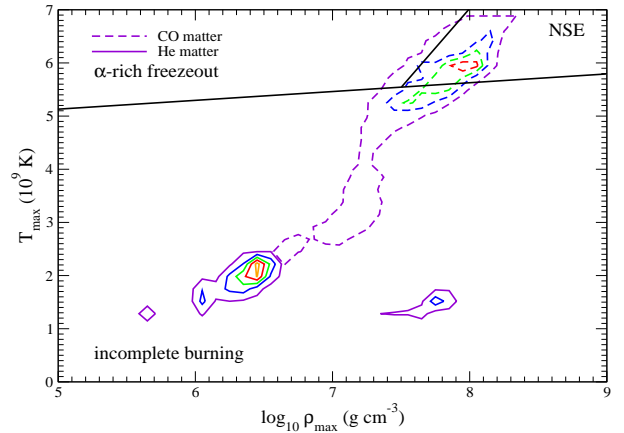}
\caption{$T_{{\rm max}}$ against $\rho_{{\rm max}}$ 
for the sub-Ch-mass benchmark model according to 
the thermodynamic trajectories. Contours correspond to 
tracer particle numbers of 100 (purple), 300 (blue), 500 (green), 700 (red), and 
900 (orange) for the C+O matter (dashed lines), and those with 10 times smaller numbers for the He matter 
(solid lines).}
\label{fig:traj_sChand_std}
\end{figure}

As in Figure \ref{fig:traj_Chand_std}, 
Figure \ref{fig:traj_sChand_std} shows 
the distribution of $T_{{\rm max}}$ against $\rho_{{\rm max}}$ for
the sub-Ch-mass benchmark model. Due to the detonation nature, 
there is always a wide spectrum of $T_{{\rm max}}$ for 
a given $\rho_{{\rm max}}$. This means that the detonation
waves inside the stars can efficiently re-heat the matter,
even when the matter is completely burnt. Compared
to the near-Ch-mass model, this model can achieve similar 
$T_{{\rm max}}$ even with a lower $\rho_{{\rm max}}$. This is 
because part of the tracer particles 
can encounter much stronger shock heating due to
geometric convergence, especially near the center.

\begin{figure}
\centering
\includegraphics*[width=8cm,height=5.7cm]{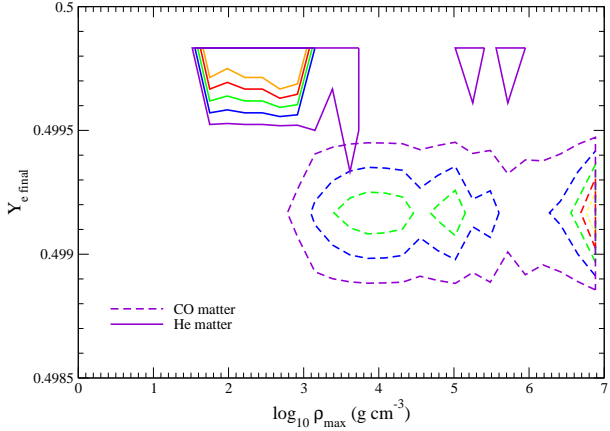}
\caption{$Y_{e,{\rm~min}}$ against $\rho_{{\rm max}}$ 
for the sub-Ch-mass benchmark model according to 
the thermodynamic trajectories. 
The contours are the same as in Figure \ref{fig:traj_sChand_std}.}
\label{fig:ye_sChand_std}
\end{figure}

As in Figure \ref{fig:ye_Chand_std},
Figure \ref{fig:ye_sChand_std} shows the
distribution of $Y_e$.
Compared to the near-Ch-mass counterpart,
there are much less tracer particles where
significant electron capture takes place.
Although the maximum $\rho_{{\rm max}}$
can be comparable to the near-Ch-mass model,
the high density is due to shock compression and
the time duration for the particle to remain
in such a density is comparatively short. Therefore, 
the fluid elements have less time 
to carry out weak interactions than
in the near-Ch-mass model.
Therefore, only a few particles can be found at relatively low $Y_e$ as $\sim 0.499$.
Note that the range of $Y_e$ is much smaller than in Fig. \ref{fig:ye_Chand_std}.

The nucleosynthesis yields, [$X_i/^{56}$Fe], are shown
in Figure \ref{fig:final_sChand_std} 
for the sub-Ch
mass benchmark model (solid line),
comparing to the model with a thicker helium envelope.
The star is completely burnt, and only small amounts of 
C, O, and Ne are left. Intermediate
mass elements from Si to Ca show the ratios close to half-solar
values.
With $M({\rm He})=0.1M_\odot$ (dashed line),
there is a large enhancement of $^{48}$Ti, $^{51}$V, and $^{52}$Cr.
This is related to 
the helium detonation, especially during the end of 
He detonation. The iron-peak elements are also 
healthily produced, except for Mn.

\begin{figure}
\centering
\includegraphics*[width=8cm,height=5.7cm]{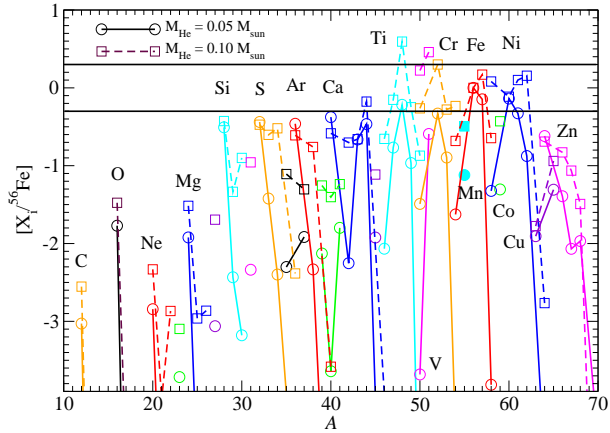}
\caption{$[X_i/^{56}{\rm Fe}]$ of stable isotopes 
in the sub-Ch-mass benchmark model (solid lines)
after short-lived radioactive isotopes have 
decayed. The ratios are scaled to the solar ratios. The horizontal lines at 
$\pm 0.3$ correspond to 0.5 or 2.0 times the solar values.
A similar model but with a thicker helium layer $M({\rm He})=0.1M_\odot$ (dashed lines) is shown for comparison. 
}
\label{fig:final_sChand_std}
\end{figure}

\begin{figure}
\centering
\includegraphics*[width=8cm,height=5.7cm]{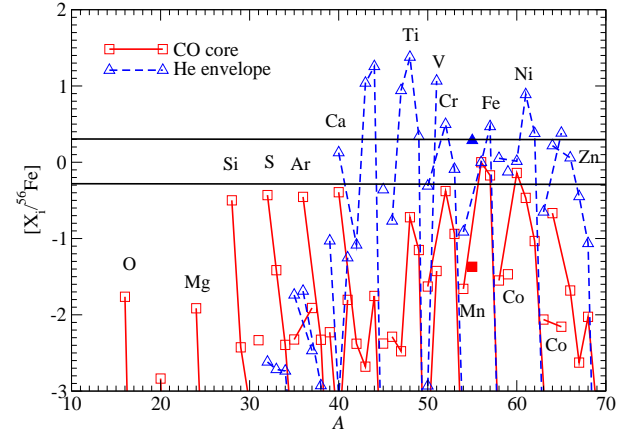}
\caption{The same as Figure \ref{fig:final_sChand_std} but for the particles ignited by carbon (solid lines) and helium (dashed lines) detonation, respectively.}
\label{fig:final_sChand_henco_std}
\end{figure}

In Figure \ref{fig:final_sChand_henco_std} we plot
[$X_i/^{56}$Fe] for the sub-Ch-mass benchmark
model with the He- and C-detonation components, separately.
Again, in the C-detonation component (solid line),
since low density matter in the core is also detonated,
small amounts of C, O, and Ne remain.
Intermediate mass elements are still produced. 
Sc, Ti, and Cr are under-produced, unlike the full abundance
profile in Fig. \ref{fig:final_sChand_std}. 
Among iron-peak elements, only $^{57}$Fe and $^{60}$Ni are sufficiently
produced. On the other hand, the He-detonation (dashed line) produces a very different abundance
pattern. Intermediate mass elements are significantly under-produced.
In contrast, there is a large enhancement of Ti, Cr, and V, with
ratios to $^{56}$Fe as large as $\sim 30$ times
the solar values.
Iron-peak elements from Mn to Zn 
look enhanced, but this is due to
the small production of $^{56}$Ni.
Note that the mass of the He envelope is 20 times smaller than that of the C+O core. %M(56Ni) = 0.01 vs 0.6

\begin{figure}
\centering
\includegraphics*[width=8.5cm]{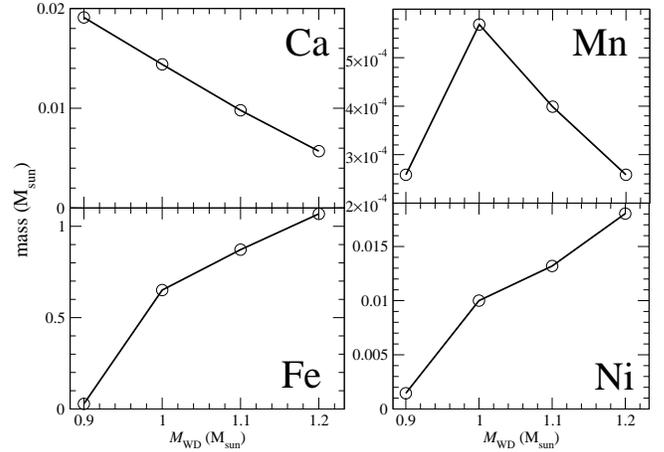}
\caption{The total masses of Ca, Mn,
Fe, and Ni in the ejecta of sub-Ch-mass
models as a function of WD mass. All models have the initial
metallicity $Z = 0.02$.}
\label{fig:element_mass}
\end{figure}

In Figure \ref{fig:element_mass} we plot the 
total yielded mass of Ca, Mn, Fe and Ni in the ejecta for
our sub-Ch-mass models as a function of WD mass. 
Clear trends can be observed for all elements. 
The mass yields of Fe and Ni are monotonically 
increasing against WD, while that of Ca is monotonically decreasing.
In contrast,
Mn mass increases and then decreases with a 
transition at $M_{\rm WD} = 1.0 M_{\odot}$. These trends
show how the C-detonation strength contributes
to the formation and destruction of elements
during nucleosynthesis. For the intermediate mass
elements such as Si, S, and Ca, when the WD mass increases
the C+O-fuel is more likely to undergo complete
burning until NSE, and thus the nuclear reactions
do not stop at Ca but continue to form iron-peak elements.
This also explains the monotonic increase in Fe and Ni with WD mass.
The falling part of Mn is also a consequence
of the strong C-detonation, which gives 
more NSE-burning instead of $\alpha$-rich freezeout.
The rising part of Mn is caused by suppression of
the incomplete and complete Si-burning
at the globally low density in 
low mass WDs.

\subsection{Comparison between Benchmark Models}

\begin{figure}
\centering
\includegraphics*[width=8.5cm]{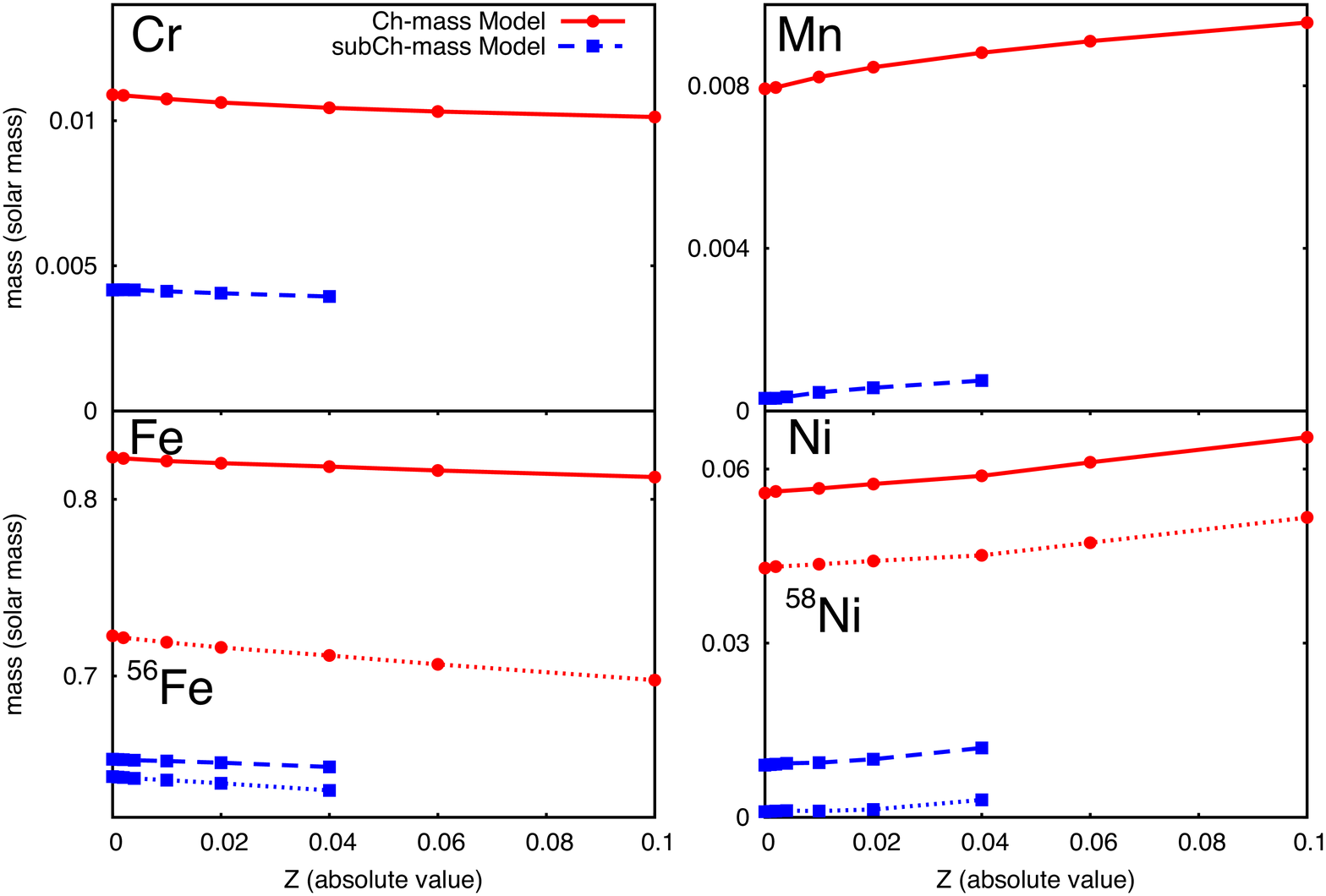}
\caption{The total masses of Cr, Mn,
Fe, and Ni
as a function of metallicity for the Ch (red solid lines)
and sub-Ch (blue dashed lines) mass benchmark models. 
The dotted lines show the masses of 
the major isotopes, $^{56}$Fe and $^{58}$Ni.
}
\label{fig:comp_element_plot}
\end{figure}

\begin{figure}
\centering
\includegraphics*[width=8.5cm]{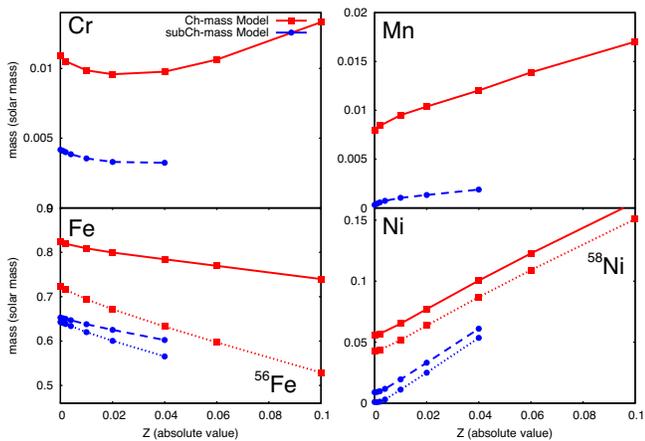}
\caption{The same as Fig. \ref{fig:comp_element_plot} but for the LN19 yields with only $^{22}$Ne for the $Z$ component of the initial composition.}
\label{fig:comp_element_plot_Ne}
\end{figure}

Finally, in Figure \ref{fig:comp_element_plot}, we plot the total yielded
masses of Cr, Mn, Fe, and Ni in the Ch
and sub-Ch mass models as a function of 
initial metallicity $Z$.
The metallicity dependence is significantly different from the yields in LN19, which are shown in Figure \ref{fig:comp_element_plot_Ne} for comparison.
In general, Cr, Mn, Ni are produced more in near-Ch-mass models than in sub-Ch-mass models by a factor of $\sim 2,10,6$, respectively, and the metallicity dependence for Mn and Ni is stronger (i.e., Z=0 to Z=0.04) in sub-Ch-mass models than in near-Ch-mass models.

The total Cr mass decreases when 
$Z$ increases. 
This trend comes from the 
the lower energy releases with higher $Z$.
At $Z > 0.04$ in Figure \ref{fig:comp_element_plot_Ne}, however, Cr mass increases with $Z$.
For these near-Ch-mass models, when $Z$ further increases, 
deflagration is further suppressed, leaving 
more matter to be burnt by detonation.
Note that Cr is produced not only by deflagration and but also by detonation (Fig. \ref{fig:final_Chand_defndet_std}).

The total Mn mass increases monotonically with $Z$
because the initial $^{22}$Ne is the seed of $^{55}$Mn.
Mn is much more produced in near-Ch-mass models than in sub-Ch-mass models.
This is due to electron capture during the initial deflagration phase, 
where more matter can have the $Y_{\rm e}$ required to form Mn.
In near-Ch-mass models, Mn is mainly produced by NSE during deflagration via $^{52}$Fe($\alpha$,p)$^{55}$Co, and a ten times smaller amount of Mn can also be produced by incomplete Si-burning at detonation (Fig. \ref{fig:final_Chand_defndet_std}) depending on $Z$. %0.02 vs 0.002
In sub-Ch-mass models, Mn mostly comes from incomplete Si-burning at He detonation (Fig. \ref{fig:final_sChand_henco_std}), which also depends on $Z$.

The total Fe mass decreases monotonically with $Z$
because most Fe comes from 
$^{56}$Fe, most of which comes from decay of
$^{56}$Ni (which has $Y_e = 0.5$). 
This isotope is 
produced by the ash in detonation which enters
the NSE region.
Increasing metallicity lowers the 
original $Y_e$ of the fuel. As
a result, even without significant electron capture 
compared to the deflagration ash, the high metallicity
automatically suppresses production of $^{56}$Ni,
and hence decreases total Fe mass. 

The total $^{58}$Ni mass increases monotonically with $Z$, 
because the 
initial $^{22}$Ne is connected to $^{58}$Ni directly
by an $\alpha$-chain (e.g., $^{54}$Fe($\alpha, \gamma$)$^{58}$Ni). 
However, this trend becomes much weaker if we adopt the solar composition for the initial metallicity (Fig. \ref{fig:comp_element_plot}).
Higher metallicity models have a 
slightly stronger detonation, which also enhances
$^{58}$Ni production.
$^{58}$Ni is produced in NSE by the deflagration in near-Ch-mass models (Fig. \ref{fig:final_Chand_defndet_std}) independent of $Z$, and also by incomplete Si-burning at detonation in near-Ch and sub-Ch mass models, depending on $Z$ (Fig. \ref{fig:final_sChand_henco_std}).

\section{Galactic Chemical Evolution}

\begin{figure}
\includegraphics[width=8.5cm]{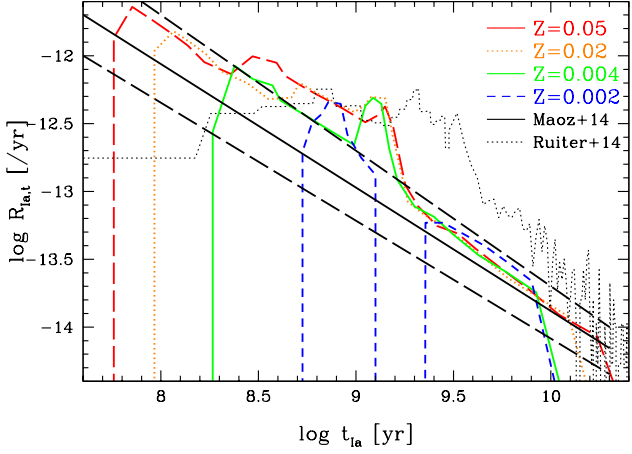}
\caption{\label{fig:ia}
Delay-time/lifetime distributions in our model, comparing to the observation from \citet[][black solid and long-dashed lines]{mao14}, and to the binary population synthesis from \citet[][black dotted line]{rui14} where all possible SN Ia progenitors are included; single-degenerate, double-degenerate, and He delayed-detonation.
}
\end{figure}

\begin{figure*}
\center 
\includegraphics[width=15cm]{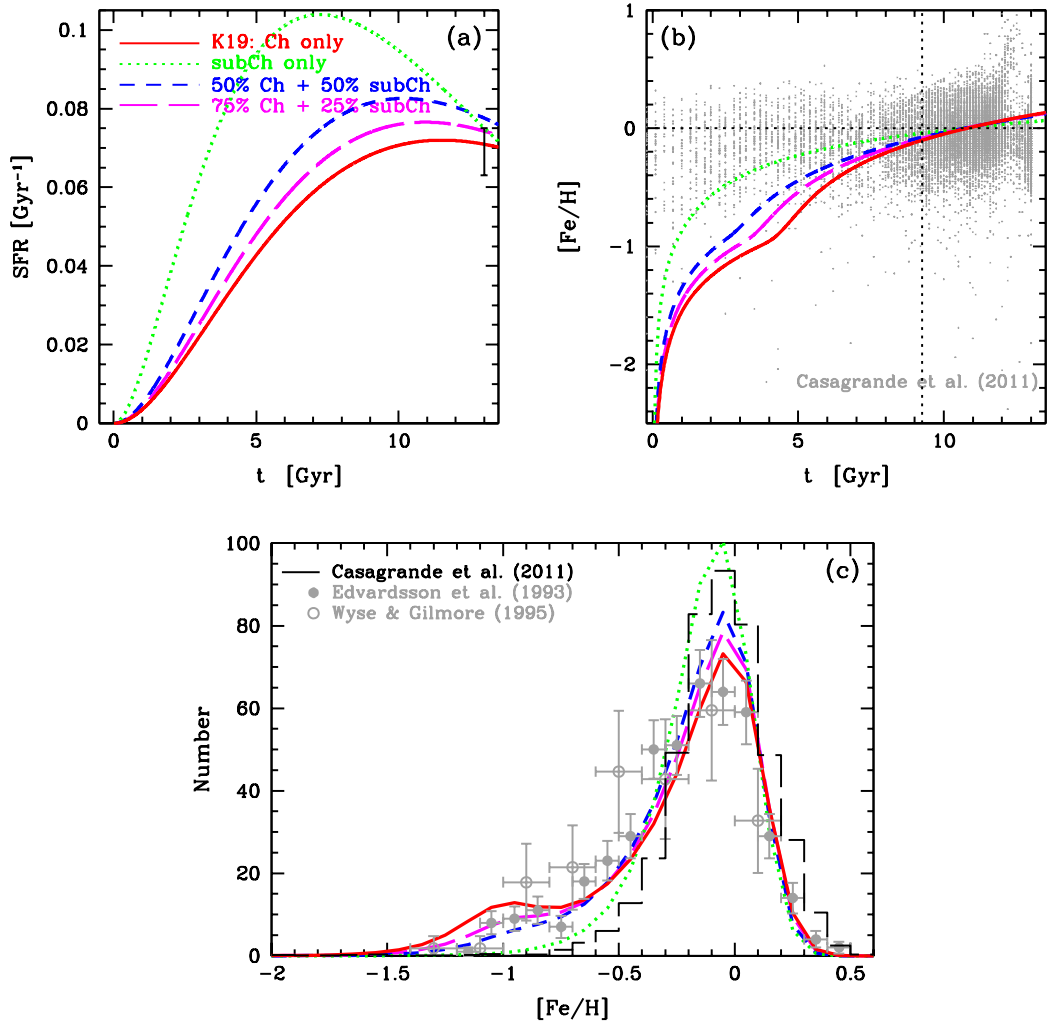}
\caption{\label{fig:mdf}
(a) Star formation history, (b) iron abundance evolution, and (c) metallicity distribution functions in the solar neighborhood for the model with Ch-mass SNe Ia only (red solid lines), sub-Ch-mass SNe Ia only (green dotted lines), 50 \% Ch and 50 \% sub-Ch mass SNe Ia (blue short-dashed lines), and 75 \% Ch and 25 \% sub-Ch mass SNe Ia (magenta long-dashed lines).
See K19 for the observational data sources.
}
\end{figure*}

\begin{figure}
\center 
\includegraphics[width=8.5cm]{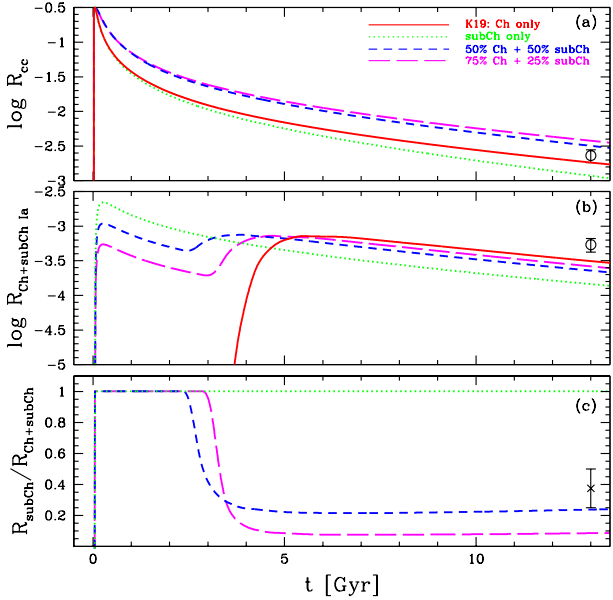}
\caption{\label{fig:snr}
The same as Fig. \ref{fig:mdf} but for supernova rate histories: the time evolution of core-collapse supernova rates (panel a), total SN Ia rate (panel b), and the ratio of sub-Ch-mass SNe Ia to the total SNe Ia (panel c).
The open circles indicate the observed SN Ia rate in a Milky Way-type galaxy taken from \citet{li11}.
The cross shows the observational estimate of the sub-Ch-mass fraction by \citet{sca14}.
}
\end{figure}

\begin{figure}[t]
\center 
\includegraphics[width=8.2cm]{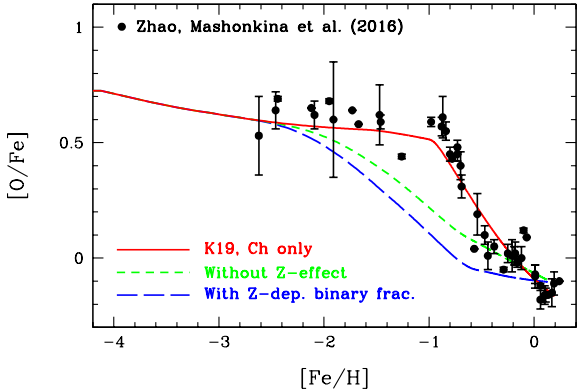}
\caption{\label{fig:iaz}
The [O/Fe]--[Fe/H] relations in the solar neighborhood for the models with Ch-mass SNe Ia only.
The red solid line is our fiducial model with the metallicity effect of WD winds. The green short-dashed line does not include the metallicity effect. The blue long-dashed line does not include it either, but include the metallicity dependence of the binary fraction from \citep{moe19}.
}
\end{figure}

\begin{figure*}
\center 
\includegraphics[width=15cm]{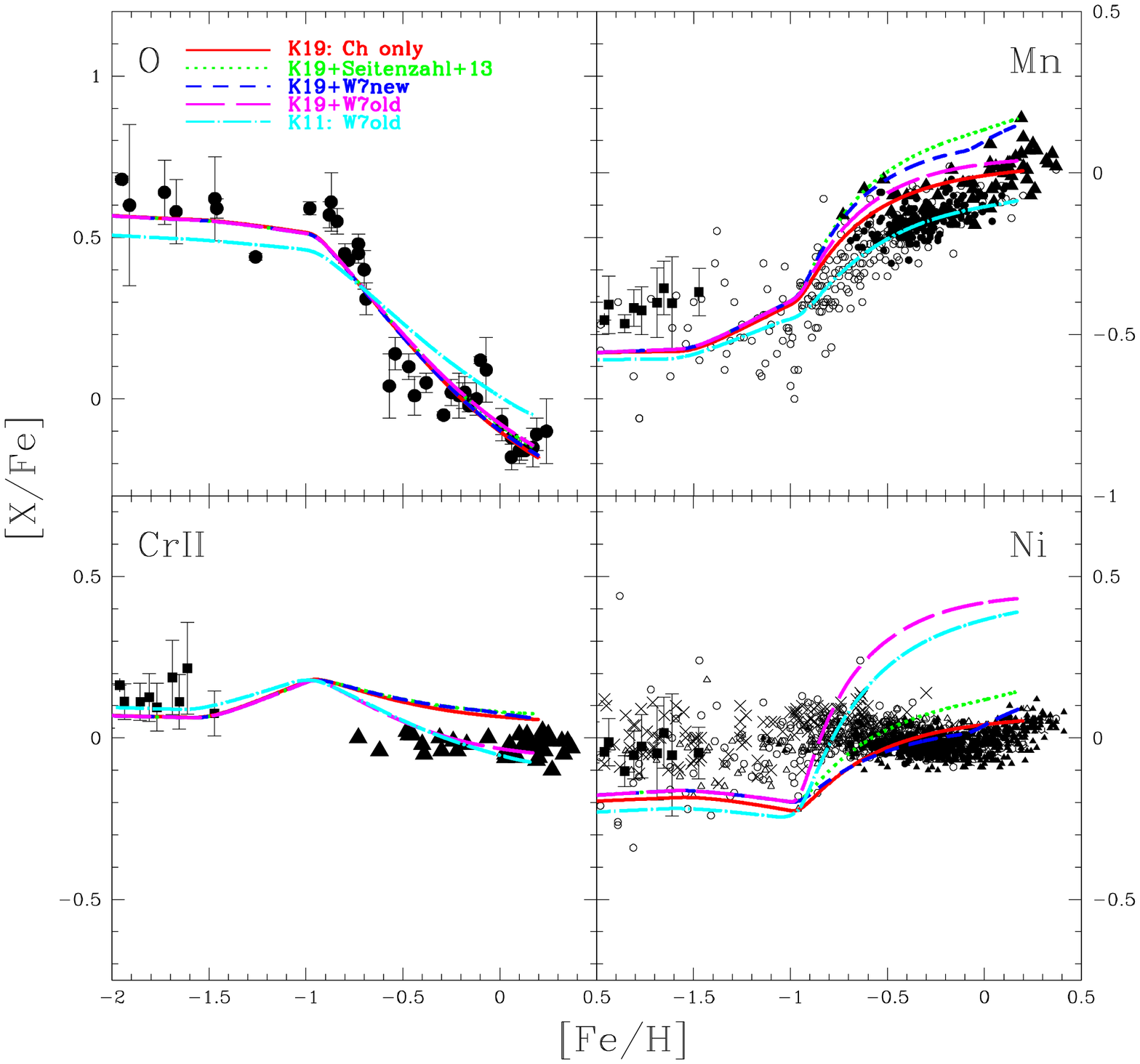}
\caption{\label{fig:ch}
Evolution of iron-peak elements against [Fe/H]
for the fiducial model with our 2D DDT yields (red solid lines), 3D DDT yields from \citet{Seitenzahl2013} (green dotted lines), the updated W7 yields (blue short-dashed lines), and the old W7 yields \citep{Nomoto1997,iwa99} (magenta long-dashed lines). The cyan dot-dashed lines are for the model in K11 where the old W7 yields are adopted.
Observational data sources are:
filled circles with errorbars, \citet{zha16};
filled squares with errorbars, \citet{reg17};
crosses, \citet{ful00};
small filled and open circles, \citet{red03,red06,red08} for thin and thick disk/halo stars;
for Cr II, 
filled triangles, \citet{ben03};
for Mn,
filled triangles, \citet{fel07};
for Ni,
small filled and open triangles, \citet{ben14} for thin and thick disk stars.
}
\end{figure*}

\begin{figure*}
\center 
\includegraphics[width=15cm]{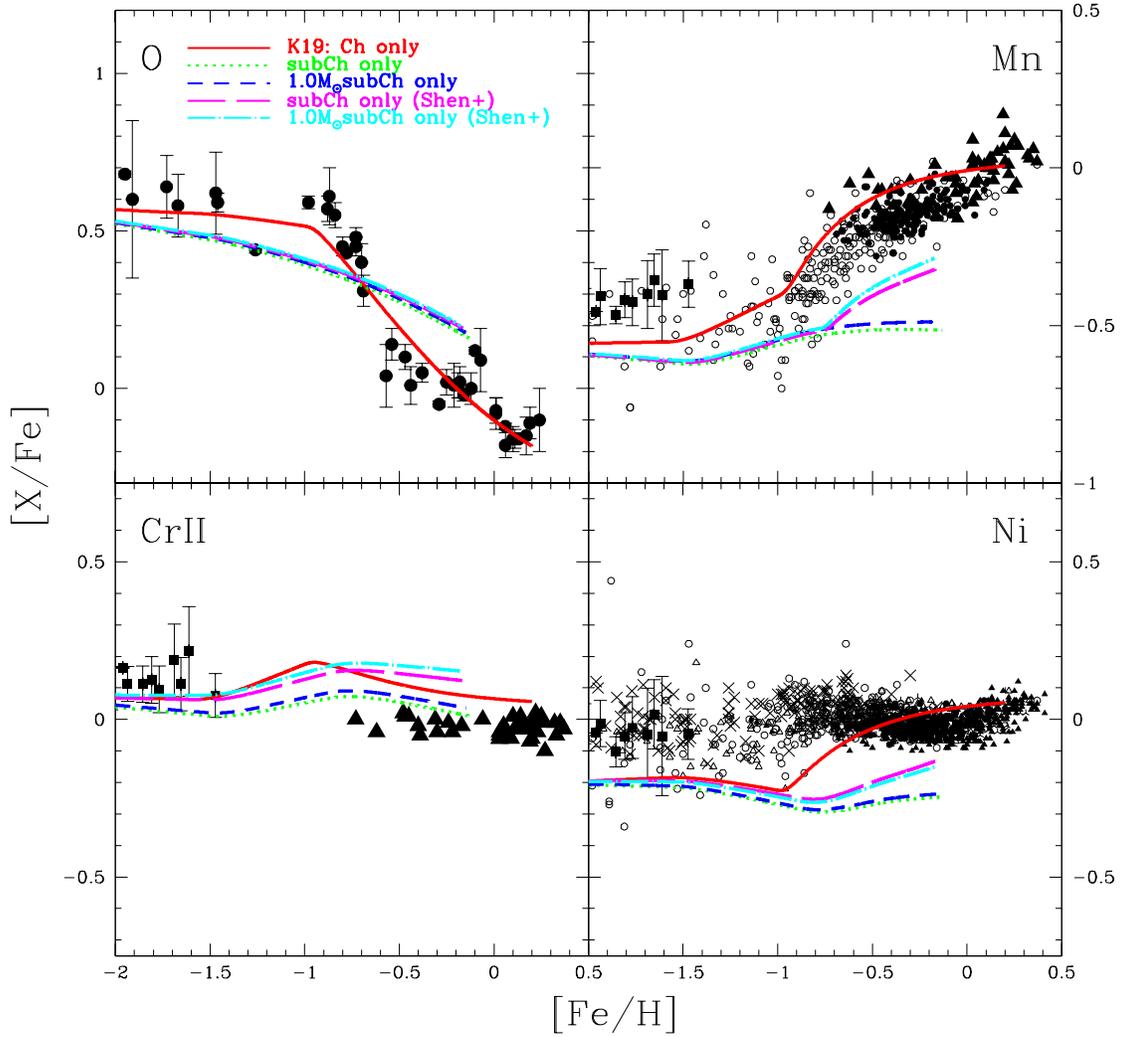}
\caption{\label{fig:subch}
The same as Fig. \ref{fig:ch} but for our 2D yields of sub-Ch-mass SNe Ia from 0.9-1.2 $M_\odot$ (green dotted lines) and $1.0 M_\odot$ (blue short-dashed lines) WDs, and 0.8-1.1 $M_\odot$ (magenta long-dashed lines) and $1.0 M_\odot$ (cyan dot-dashed lines) 1D yields from \citet{Shen2018}.
}
\end{figure*}

\begin{figure*}
\center 
\includegraphics[width=15cm]{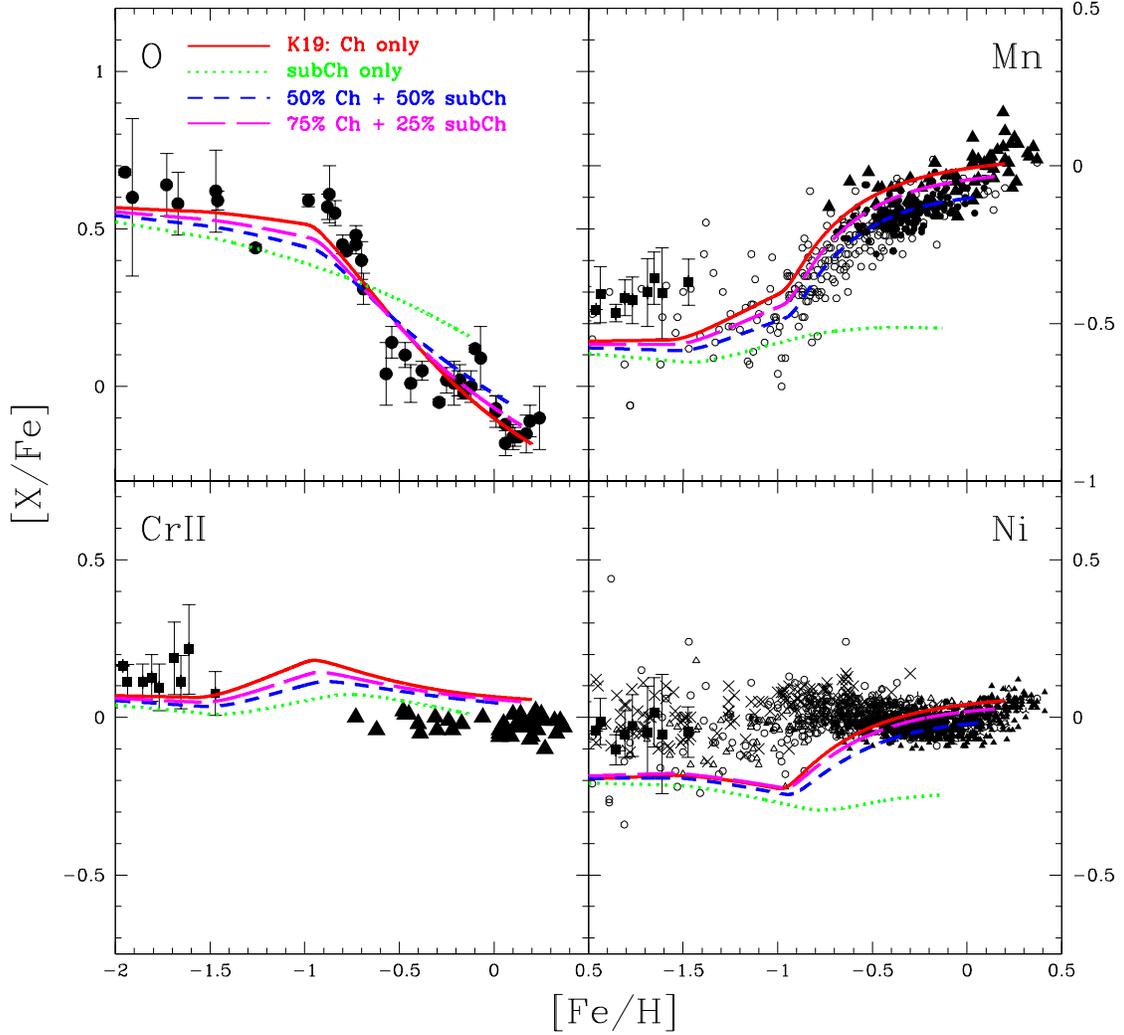}
\caption{\label{fig:both}
The same as Fig. \ref{fig:ch} but for sub-Ch-mass SNe Ia only (green dotted lines), 50 \% Ch and 50 \% sub-Ch mass SNe Ia (blue short-dashed lines), and 75 \% Ch and 25 \% sub-Ch mass SNe Ia (magenta long-dashed lines).
}
\end{figure*}

\begin{table}
\center
\caption{\label{tab:param}
Parameters of the GCE models for the solar neighbourhood (SN) and dwarf spheroidal galaxies: timescales of infall ($\tau_{\rm i}$), star formation ($\tau_{\rm s}$), and outflow ($\tau_{\rm o}$), and the galactic wind epoch $\tau_{\rm w}$, all in Gyr.}
\begin{tabular}{l|cccc}
\hline
 & $\tau_{\rm i}$ & $\tau_{\rm s}$ & $\tau_{\rm o}$ & $t_{\rm w}$ \\
\hline
SN, Ch only     & 5 & 4.7 & - & - \\
SN, 75\% sub-Ch & 5 & 4.0 & - & - \\
SN, 50\% sub-Ch & 5 & 3.2 & - & - \\
SN, sub-Ch only & 5 & 1.0 & - & - \\
\hline
Fornax              & 10 ($t<9$) & 25 & 5 & 12 \\ 
Fornax, no sub-Ch   & 10 ($t<9$) & 15 & 5 & 12 \\ 
Sculptor            & 1 & 50 & 1 & 9 \\
Sculptor, no sub-Ch & 1 & 40 & 1 & 9 \\
Sextans             & 0.5 & 100 & 1.4 & 7 \\
Sextans, no sub-Ch  & 0.5 & 70  & 1.4 & 7 \\
Carina              & 0.1 ($t<6.5$) & 200 & 5 & 12 \\
                    & 100 ($t=6.5-9$) & & & \\
Carina, no sub-Ch   & 0.1 ($t<6.5$) & 100 & 5 & 12 \\
                    & 100 ($t=6.5-9$) & & & \\
Carina, Iax*50      & 0.1 ($t<6.5$) & 400 & 5 & 12 \\
                    & 100 ($t=6.5-9$) & & & \\
\hline
\end{tabular}
\end{table}

\subsection{The GCE code}
\label{sec:gce}

Since nucleosynthesis yields are significantly different between Ch and sub-Ch mass SNe Ia, changing the relative contribution results in different elemental abundance ratios at a given metallicity.
The evolutionary tracks of elemental abundance ratios depend on the star formation history. However, the star formation history can be tightly constrained from the other independent observations, namely, the metallicity distribution function of stars in the system considered, and in the solar neighborhood, only a small variation is possible for the evolutionary tracks of elemental abundance ratios (see Appendix A).
Therefore, the elemental abundance ratios in the solar neighborhood have been used as the most stringent constraint for the nucleosynthesis yields of core-collapse supernovae \citep[e.g.,][]{tim95,kob06,rom10} and for the progenitor models of SNe Ia \citep[e.g.,][]{mat86,kob98}.

The evolutionary tracks of elemental abundance ratios are calculated with GCE models \citep{tin80,pagel1997,matteucci2001}, and our basic equations are described in \citet{kob00}.
The code follows the time evolution of elemental and isotopic abundances in a system where the interstellar medium (ISM) is instantaneously well mixed (and thus it is also called a ``one-zone'' model).
 No instantaneous recycling approximation is adopted and thus chemical enrichment sources with long time-delays such as SNe Ia are properly included.

The stellar physics/empirical relations included in our GCE models are as follows;
The star formation rate (SFR) is proportional to the gas fraction, which evolves with inflow and outflow to/from the system considered as well as mass-loss from dying stars and supernova explosions.
The Kroupa initial mass function (IMF) is adopted \citep[][hereafter K11]{kob11agb}. The nucleosynthesis yields of core-collapse supernovae, Type II supernovae and hypernovae (HNe), are also taken from K11\footnote{The yield table is identical to that in \citet{Nomoto2013}.} but with failed supernovae at $>30M_\odot$ (K19). The HN fraction depends on the metallicity; $\epsilon_{\rm HN}=0.5, 0.5, 0.4, 0.01$, and $0.01$ for $Z=0, 0.001, 0.004, 0.02$, and $0.05$ \citep{kob11mw}.
Then, the gas infall and star formation timescales, $\tau_{\rm i}$ and $\tau_{\rm s}$, are determined to match the observed metallicity distribution function (MDF) of the system.
As shown in Appendix A, the set of $\tau_{\rm i}$ and $\tau_{\rm s}$ can be uniquely determined from the MDF (Fig.\ref{fig:gce2}), and we chose $\tau_{\rm i}=5$ and $\tau_{\rm s}=4.7$ for our fiducial model (K11). Our conclusions are not affected by this choice of these parameters.

\subsection{SN Ia model}
\label{sec:snia}
For Ch-mass SNe Ia ($1.37M_\odot$), we introduced our formulation for the rate in K98, and the details are discussed in \citet[][hereafter KN09]{kob09}.
The lifetime distribution function (also called delay-time distribution, DTD) is calculated 
with Eq.[2] in KN09; it multiplies the mass functions of primary and secondary stars, and gives a very similar results to the formula in \citet{gre83}.
The main difference is that we include the metallicity dependence of secondary mass ranges due to the requirements of the WD optically-thick winds \citep{kob98} and
the mass-stripping on the binary companion stars (KN09).
As a result, our SN Ia rate drops at lower [Fe/H] than $\sim -1.1$ (i.e., $A({\rm O}) \sim 7.6$).
Without this metallicity effect on the SN Ia rate, it is not possible to reproduce the observed [O/Fe]--[Fe/H] relation in the solar neighborhood (K98, see also Fig.\ref{fig:iaz}).
This metallicity cut is not inconsistent with the observed metallicities of host galaxies (KN09), although such metallicity dependence has not yet been seen in the observations of SN Ia rates.
In the observed mass-metallicity relation of galaxies, the stellar metallicity of [Fe/H] $\sim -1.1$ corresponds to a stellar mass of $\sim 10^{7-8}M_\odot$ \citep{zah17}, above which the specific SN Ia rate increases toward lower-mass galaxies but is uncertain below this mass \citep{bro19}.

The range of lifetimes are determined from the mass ranges of secondary stars (K09).
As a result,
the main sequence (MS) star$+$WD systems have timescales of 
$\sim 0.1-1$ Gyr, which are dominant in star-forming galaxies and correspond to the observed ``prompt'' population \citep{man06,sul06},
while the red giant (RG) star$+$WD systems have lifetimes of $\sim 1-20$ Gyr, which are dominant in 
early-type galaxies and correspond to the ``delayed'' population.
The shortest lifetime depends on the maximum secondary mass, which depends the progenitor metallicity (KN09). The longest lifetime is determined from the minimum secondary mass, which is $0.9 M_\odot$ independent of metallicity. At high metallicity, stellar luminosity is lower due to the higher opacity, which results in a longer stellar lifetime. Therefore, our SN Ia lifetime becomes as long as 20 Gyr at solar metallicity. This metallicity dependent lifetime was not taken into account in \citet{hac08}.

Finally, the normalization, i.e., the absolute rate of (Ch-mass) SNe Ia, is determined by two binary parameters respectively for the MS+WD and RG+WD systems, $b_{\rm MS}$ and $b_{\rm RG}$.
The set of $b_{\rm MS}$ and $b_{\rm RG}$ can be uniquely determined from the [O/Fe]--[Fe/H] relation at [Fe/H] $>-1$ (Fig.\ref{fig:gce3}), and we chose $b_{\rm RG}=0.02$ and $b_{\rm MS}=0.04$ for our fiducial model.
This choice does not affect our conclusions either, as shown in Figure \ref{fig:gce3} of Appendix A.
The binary parameters include not only binary fractions, but also separations and any other conditions that successfully lead SN Ia explosions, and the numbers are defined as the fractions of WDs that eventually explode as SNe Ia at $Z=0.004$.
At $Z \sim 0.02$, the resultant delay-time/lifetime distribution is very similar to that derived from observed supernova rates in nearby galaxies \citep{mao14}.
Figure \ref{fig:ia} shows the comparison of observational and theoretical distributions to ours.

For sub-Ch-mass SNe Ia, we used the same formula for single degenerate systems in \citet{kob15}. In this paper, however, in order to include sub-Ch-mass SNe Ia both from single and double degenerate systems, we use the ``observed'' delay-time distribution function instead; the rate is $10^{-13} {\rm yr}^{-1} M_\odot^{-1}$ at 1 Gyr, proportional to $1/t$ over 0.04-20 Gyr, independent of metallicity.
For single degenerate systems, the rate could be higher for lower metallicities because the maximum secondary mass becomes lower (see Fig.1 of \citealt{kob15}).
For double degenerate systems, however, no such a metallicity dependence is expected, but this should be studied in more details with binary population synthesis.
For sub-Ch-mass SNe Ia, the nucleosynthesis yields not only depend on metallicity but also depend on the masses of the primary WD.
Since the mass distribution of primary WDs is uncertain, we add the contributions from $0.9, 1.0, 1.1$, and $1.2M_\odot$ WDs respectively with 10\%, 40\%, 40\%, and 10\% for our 2D double detonation yields.
With this weighting, on average sub-Ch-mass SNe Ia give less Fe mass per event than for Ch-mass SNe Ia.

\subsection{Star Formation History}
In this paper, we show four GCE models with Ch-mass SNe Ia only (solid lines), sub-Ch-mass SNe Ia only (dotted lines), 50\% each contribution (short-dashed lines), and 75\% Ch-mass and 25\% sub-Ch-mass SNe Ia contributions (long-dashed lines).
For each case, star formation and infall timescales are chosen to reproduce the observed metallicity distribution function (Fig. \ref{fig:mdf}c), as well as the present SFR (Fig. \ref{fig:mdf}a) and the solar metallicity at 4.6 Gyr ago (Fig. \ref{fig:mdf}b).
In the models with sub-Ch-mass SNe Ia, because of lower Fe production (on average), higher SFRs are required to obtain the same peak metallicity of the MDF, compared with the model with Ch-mass SNe Ia only (Table \ref{tab:param}).

Figure \ref{fig:snr} shows supernova rate histories.
Reflecting the small difference in the SFRs, there is also a small difference in the core-collapse supernova rates (panel a).
The middle panel shows the total SN Ia rates.
With the ``observed'' delay-time distribution, sub-Ch-mass SNe Ia start to occur $0.04$ Gyr after the onset of galaxy formation, and the rate monotonically decreases as a function of time.
Therefore, with a larger fraction of sub-Ch-mass SNe Ia, the total SN Ia rate becomes higher at early epochs.
This means that the ISM reaches [Fe/H] $=-1.1$ earlier, Ch-mass SNe Ia start to occur earlier with our progenitor model, and thus the second peak caused by Ch-mass SNe Ia also appears earlier.
The model with Ch-mass SNe Ia only gives the best match to the observed rate for Milky-Way type galaxies.
Note that, however, this is not a totally fair comparison since the observed values are for the entire galaxy while the model is for the solar neighborhood.

It is difficult to observationally estimate the sub-Ch-mass fraction in the total SN Ia rate; it requires estimating ejecta mass from supernova light curves modelling as well as handling the selection bias of observed supernovae.
\citet{sca14} estimated the sub-Ch-mass fraction at 25-50\% in their unbiased sample of spectroscopically normal SNe Ia.
The bottom panel shows the fraction of sub-Ch-mass SNe Ia to the total SNe Ia rate for our models, which evolves as a function of time. It is 100\% at the beginning, while it decreases once Ch-mass SNe Ia starts to occur.
At present, 50\% GCE contribution of sub-Ch-mass SNe Ia results in 25\% of sub-Ch-mass fraction in the SN Ia rate (blue short-dashed line), which is in reasonable agreement with the observational estimate.
This fraction also depends on the evolutionary phase of the galaxy, and hence on the type/mass of the host galaxies.
On the other hand, the observational estimate of the sub-Ch-mass fraction is for the average of various types/masses of galaxies with various stellar ages.
Because of these reasons, we do not adopt the ``observed'' sub-Ch-mass fraction, but instead aim to constrain the fraction using GCE modes.

Figure \ref{fig:iaz} shows the resultant [O/Fe]--[Fe/H] relations for the models with Ch-mass SNe Ia only.
Without the metallicity effect on the WD winds, it is not possible to reproduce the observed evolutionary change of [O/Fe] at [Fe/H] $\sim -1$ (see \S \ref{sec:xfe} for the observational data).
As noted above, changing star formation timescale would not solve this problem, while reproducing the observed MDF (see also Fig.\ref{fig:gce2}).
Recently, metallicity dependence of the binary fraction is indicated from observations \citep{moe19}, where the binary fraction is higher at lower metallicities.
If we scale our binary parameter ($b_{\rm MS}$ and $b_{\rm RG}$) to the observed metallicity-dependent binary fraction, then there are many more SNe Ia at earlier epochs, which decreases the [O/Fe] ratios even further away from the observational data. 
Therefore, it is necessary to include our metallicity effect of Ch-mass SNe Ia in order to reproduce this most important observation of GCE.

\subsection{Elemental Abundance Ratios}
\label{sec:xfe}

Not only [O/Fe] ratios but also abundance ratios among iron-peak elements are the key to constrain the fraction of sub-Ch-mass SNe Ia.
For the elemental abundance ratios of individual stars, the most accurate observational data, i.e., high-resolution observations with star-by-star analysis, are available for the solar neighborhood.
We take the non-local-thermodynamic equilibrium (NLTE) abundances for oxygen \citep{zha16}, while LTE abundances are used for iron-peak elements \citep[e.g.,][]{red03,fel07,ben14,reg17}.
The NLTE effects of iron-peak abundances could also be important. It is worth noting, however, that the effects may not be so large with the updated atomic data (\citealt{sne16}, but see \citealt{ber08}).
The exception is for Cr, and we plot Cr II observations (see \citealt{kob06} for the comparison between Cr I and Cr II observations).

Figure \ref{fig:ch} shows the evolution of elemental abundances against [Fe/H] for the models with various yields of Ch-mass SNe Ia.
[O/Fe] shows a decrease from [Fe/H] $\sim -1$ to higher [Fe/H], while [Mn/Fe] shows an increase; these opposite behaviours are well reproduced by the delayed enrichment of SNe Ia.
The observed [Ni/Fe] ratios show a constant value of $\sim 0$ over all the metallicity range.
It has been known that the W7 yields \citep{Nomoto1997,iwa99} over-produce Ni by $\sim 0.5$ dex (magenta long-dashed and cyan dot-dashed lines; see also Fig. 24 of \citealt{kob06}).
This Ni over-production problem is mostly solved with the updated nuclear reaction rates, mainly due to the lower electron-capture rates (blue dashed lines)\footnote{The updated W7 yields were presented in \citet{Nomoto2018} and LN18, which are re-calculated with new initial composition and decay time in this paper.}.
$Y_{\rm e}$ becomes higher approaching to 0.5, which gives lower [(Ni, Co)/Fe] and higher [(Cr, Mn)/Fe].
Our 2D DDT yields of $1.37M_\odot$ give very similar results (red solid lines) as the updated W7 yields, but the [Mn/Fe] ratio is reduced by 0.1 dex because of slower flame speed in our more realistic 2D model, which gives a better agreement with observations.
The 3D DDT yields from \citet{Seitenzahl2013} also give very similar results (green dotted lines) as our 2D DDT yields but with 0.1 dex higher [(Mn, Ni)/Fe] ratios. This is probably because their multi-ignitions result in more material to be burnt by deflagration waves.

In summary, with the updated electron-capture rates, these three models (W7, 2D DDT, and 3D DDT) of Ch-mass SNe Ia give the elemental abundance ratios within the observational scatters in the solar neighborhood, and our 2D DDT gives the best fit to [Mn/Ni] ratios at $-1 \ltsim$ [Fe/H] $\ltsim 0.3$.
The [Ni/Fe] ratio still shows a mild increase from $-0.2$ to $+0.05$ with [Fe/H]. Whether this is inconsistent or not depends on the yields of core-collapse supernovae that determine the plateau value of [Ni/Fe] at [Fe/H] $\ltsim -1$. Since both Ni and Fe are formed at the innermost regions of core-collapse supernovae, multidimensional effects can change the Ni/Fe ratios.

Although the star formation history is the same as in K11 (cyan dot-dashed lines) and K19 (red solid lines), the K19 model is updated by including faint supernovae, a metal-dependent HN fraction, and different SN Ia parameters.
These result in a slightly higher time-integrated SN Ia rate with slightly later start (see Fig.\ref{fig:snr}), which leads to the lower [O/Fe] and higher [(Mn, Ni)/Fe] ratios than in the K11 model, at [Fe/H] $\sim 0$.
The better match of Ni is not due to the GCE modelling; the K19 model with the old W7 yields (magenta long-dashed lines) still shows the over-production of Ni.

Figure \ref{fig:subch} shows the impact of sub-Ch-mass SNe Ia on GCE models.
There are two problems when compared with observations.
First of all, the decrease of Fe yield results in a much shallower [O/Fe] slope. The difference around [Fe/H] $\sim -1$ is caused by the metal-independent delay-time distribution adopted for sub-Ch-mass SNe Ia; sub-Ch-mass SNe Ia occur earlier than Ch-mass SNe Ia in the solar neighborhood models.
Secondly, the [Mn/Fe] ratio is much lower, giving almost no increase at [Fe/H] $\gtsim -1$, which cannot explain the observations.
For Ni, if the Ni/Fe ratio of core-collapse supernovae becomes 0.2 dex higher, sub-Ch-mass SNe Ia might not be inconsistent with the observations.
The [Cr/Fe] is 0.05 dex lower than the Ch-mass model, which is more consistent with the observations.

As noted in \S \ref{sec:snia}, there is a WD mass dependence on the yields of sub-Ch-mass SNe Ia, and $1.0M_\odot$ WD models gives relatively high Mn/Fe ratios (Fig. \ref{fig:element_mass}).
Even if we include $1.0M_\odot$ WDs only (blue short-dashed lines), our conclusion is unchanged; only with sub-Ch-mass SNe Ia, it is impossible to produce enough Mn to explain the observations in the solar neighborhood.

Our 2D yields are notably different from the 1D detonation yields from \citet{Shen2018} (magenta long-dashed lines), with which [(Cr, Mn, Ni)/Fe] ratios are higher than in our models.
Here we add the contributions from $0.8, 0.85, 0.9, 1.0$, and $1.1 M_\odot$ WDs respectively with 5\%, 5\%, 10\%, 40\%, and 40\%.
The difference is not caused by this summation; for the model with $1.0M_\odot$ only (cyan dot-dashed lines), we also find the same difference between our yields and \citet{Shen2018}'s yields.
Note that their yields were calculated using their 1D code without any He layer, and with the initial metallicity given by $^{22}$Ne only. Although a larger network is included during hydrodynamical calculation, the network during post-processing is smaller than ours.

In summary, compared with observations, none of these sub-Ch-mass SN Ia models (1D and 2D) show better a match than in the Ch-mass SN Ia model (red solid lines).
Sub-Ch-mass SNe Ia may help further in solving the Ni over-production problem, at the expense of reproducing the observed [$\alpha$/Fe] and [Mn/Fe]. Then, the next question is what fraction of SNe Ia can come from sub-Ch-mass WDs?

Figure \ref{fig:both} shows the evolution of elemental abundance ratios with varying the fraction of sub-Ch-mass SNe Ia.
If 50\% of the delay-time distribution comes from sub-Ch-mass SNe Ia, the [O/Fe] slope with [Fe/H] is too shallow, although the [(Mn, Ni)/Fe] ratios are within the scatters of observational data.
With 25\% sub-Ch-mass SNe Ia and 75\% Ch-mass SNe Ia, it is possible to reasonably reproduce all observational constraints.

\section{Dwarf Spheroidal Galaxies}
\label{sec:dsph}

\begin{figure*}
\center 
\includegraphics[width=15cm]{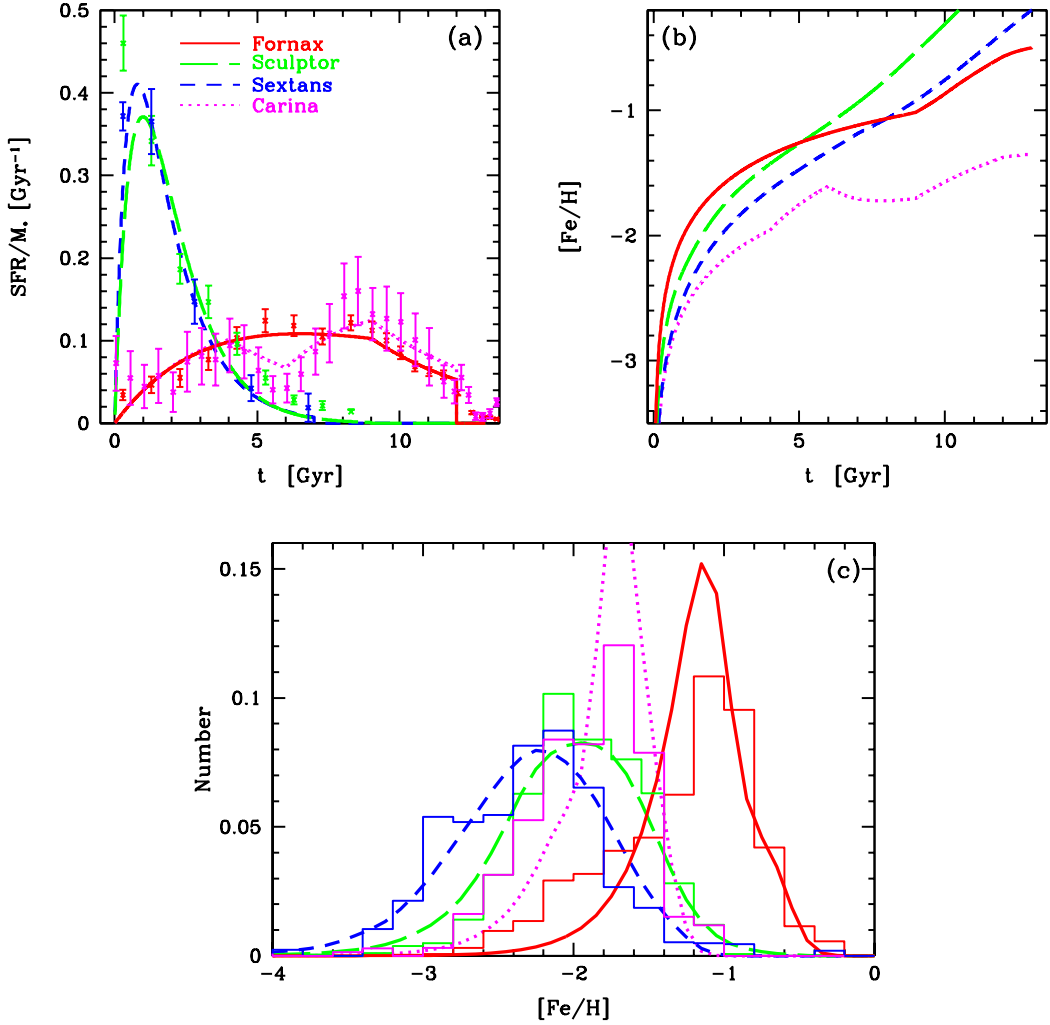}
\caption{\label{fig:dsph-mdf}
The same as Fig. \ref{fig:mdf} but for dwarf spheroidal galaxies, Fornax (red solid lines), Sculptor (green long-dashed lines), Sextans (blue short-dashed lines), and Carina (magenta dotted lines).
The observational data sources are: \citet{deboer12for}, \citet{deboer12scl}, \citet{deboer14}, \citet{lee09} for the panel (a) and \citet{sta10} for the panel (c).
}
\end{figure*}

Detailed elemental abundances are also obtained for the stars in dSph galaxies \citep[e.g.,][]{tol09}, and from the observed abundance patterns, it has been debated if dSphs galaxies are the building blocks of the Galactic halo or not. The very different abundance patterns of the stars in `classical' dSphs (with relatively large stellar masses) suggest that dSphs are not the building blocks\footnote{Metal-poor stars of ultra-faint dSph galaxies can still be the building blocks.}, but instead they provide an independent constraint on stellar physics at a different environment.

DSph galaxies are not a homogeneous population but have formed with a variety of star formation histories, and various chemical evolution models have been presented \citep[e.g.,][]{carigi02,lanfranchi06,cescutti08,vincenzo14}.
Because of the shallow potential well, the ISM can be easily blown away due to supernova feedback after the initial star burst. In addition to the description in \S \ref{sec:gce}, outflow is also included, proportional to the SFR, i.e., the gas fraction of the system, with a timescale $\tau_{\rm o}$.
If supernova energies are accumulated, star formation can be totally quenched. This is called galactic winds, and the epoch is defined with $t_{\rm w}$.
In order to constrain GCE model parameters, it is necessary to have a number of observational constraints such as MDFs, and thus we model four dSphs galaxies, Fornax, Sculptors, Sextans, and Carina, which have stellar masses of $20, 2.3, 0.44$, and $0.38 \times 10^6M_\odot$ \citep{mcc12}.
Stellar age distributions have also been estimated comparing photometric data to stellar evolutionary tracks, which are also used for constraining model parameters. The adopted parameters are summarized in Table \ref{tab:param}.

Figure \ref{fig:dsph-mdf} shows the adopted observational constraints of SFRs and MDFs, as well as the model predictions of age-metallicity relations, for the fiducial models of these four dSph galaxies. The resultant formation histories can be summarised as follows.
The models for Sculptor and Sextans are very similar; both are formed by a rapid infall and star formation with a strong outflow. Since the star formation efficiency in Sextans is lower than in Sculptor, the average [Fe/H] of the MDF is $\sim 0.3$ dex lower, which is probably due to the mass difference of the systems. 
The models for Fornax and Carina are also similar; both have extended star formation histories with longer infall timescales. Since the star formation efficiency in Carina is lower than in Fornax, the peak [Fe/H] of the MDF is $\sim 0.6$ dex lower, which is also due to the mass difference of the systems. There is also an outflow, but this is weaker than in Sculptor and Sextans.
Through dynamical interaction, it is possible to have multiple gas infalls in dSph galaxies. The observed age distribution of Carina is well reproduced by two infalls, one with a short timescale and another with a much longer timescale at time $t=6.5$ Gyr.

Because of the reasons described in the next section, in these fiducial models, the 100\% contribution of sub-Ch-mass SNe Ia is added on top of the contributions from core-collapse supernovae and Ch-mass SNe Ia\footnote{For the solar neighborhood models in \S 3, the total SN Ia rate is fixed and only the relative contribution from Ch and sub-Ch mass SNe Ia is varied.} (or SN Iax for Sculptor).
In the predicted iron abundance evolutions (panel b), chemical enrichment timescales are shorter for more massive systems. In observational data, the age--metallicity relations should have steeper slopes at $t\ltsim 3$ Gyr, although for such old stars it is very difficult to estimate age and metallicity independently.
The three dSph galaxies (except for Carina) reached [Fe/H] $\sim -1$ at $t=$ 5 to 10 Gyrs, and after that the iron abundance evolution is speeded up because of Ch-mass SNe Ia. In Carina, however, [Fe/H] never reaches $\sim -1$ as in the observed MDF (panel c), and thus there is no enrichment from Ch-mass SNe Ia\footnote{Note that, however, in more realistic hydrodynamical simulations \citep[e.g.,][]{kob11mw}, Ch-mass SNe can occur at [Fe/H] $\ltsim -1$ due to inhomogeneous enrichment, and [$\alpha$/Fe] can show a decrease in the case of strong supernova feedback.}.

\subsection{Elemental Abundance Ratios}

\begin{figure}[t]
\center 
\includegraphics[width=8.5cm]{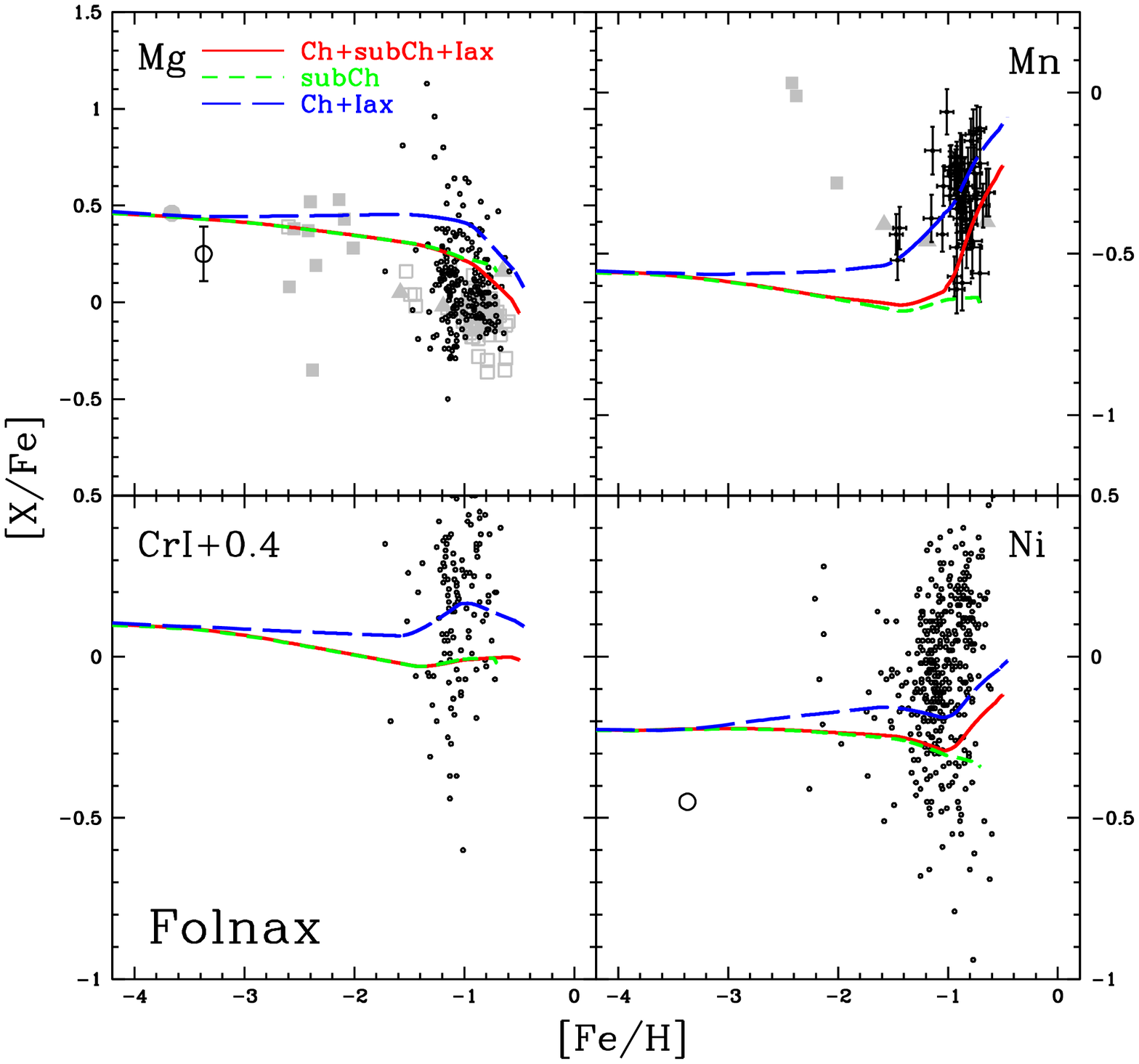}
\caption{\label{fig:fornax}
Evolution of elemental abundances against [Fe/H] for Fornax dSph galaxy with all SN Ia channels (red solid lines), sub-Ch-mass SNe Ia only (green short-dashed lines), and Ch-mass SNe Ia and SNe Iax (blue long-dashed lines).
The large points show high-resolution NLTE abundances from \citet{mas17} and LTE abundances for Mn and Cr from \citet{jab15}.
The small points show low-resolution observational data from \citet{kir19}.
The gray points show LTE abundances (see \citealt{kob15} for the references therein).
}
\end{figure}

\begin{figure}[t]
\center 
\includegraphics[width=8.5cm]{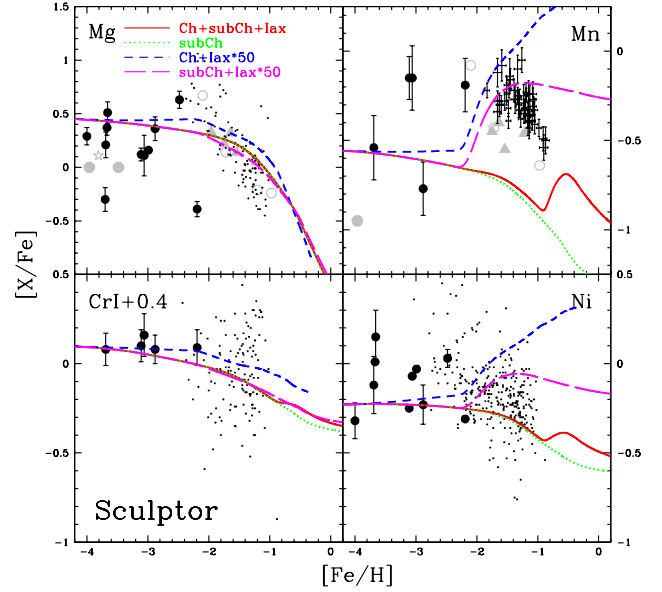}
\caption{\label{fig:sculptor}
The same as Fig.\ref{fig:fornax} but for Sculptor dSph galaxy with all SN Ia channels (red solid lines), sub-Ch-mass SNe Ia only (green dotted lines), Ch-mass SNe Ia plus 50 times more SNe Iax (blue short-dashed lines), and sub-Ch-mass SNe Ia plus 50 times more SNe Iax (magenta long-dashed lines).
}
\end{figure}

As in \S \ref{sec:xfe}, abundance ratios among iron-peak elements are the key to constrain the contribution of various types of SNe Ia in dSph galaxies.
However, it is much harder to estimate these abundance ratios because of the distance of the systems (the only observed stars are red-giants) and the limited samples of each system. In particular, NLTE analysis has been made only for a small number of stars.
For constraining models, Mg NLTE abundances and Ni LTE abundances are taken from \citet{mas17}, which uses the same NLTE model as in \citet{zha16} for the solar neighborhood.
Mn and Cr data are taken from \citet{jab15} also with their LTE analysis.
Cr abundances are obtained from the Cr I lines, which are known to underestimate Cr abundances, and a +0.4 dex shift is applied as in \citet{kob06}.
LTE Mn abundances are also taken from \citet{nor12}.
A few more stars are taken from the data compilation by \citet{ven12}, and a large sample from medium-resolution spectra by \citet{kir19}.

\begin{figure}[t]
\center 
\includegraphics[width=8.5cm]{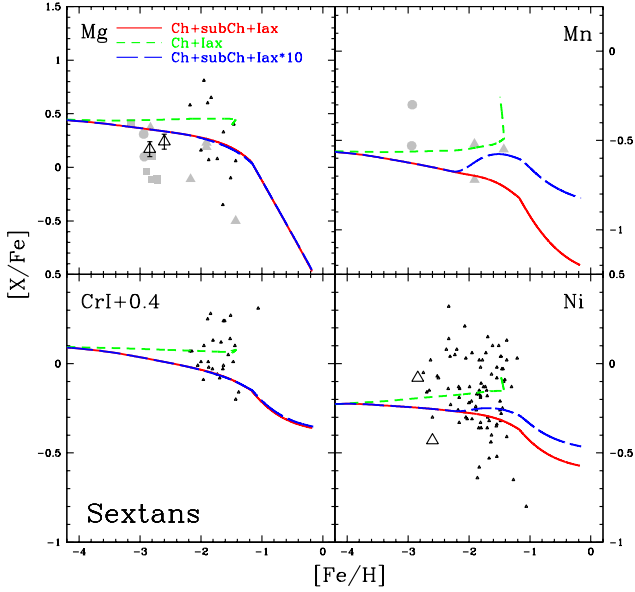}
\caption{\label{fig:sextans}
The same as Fig.\ref{fig:fornax} but for Sextans dSph galaxy with all SN Ia channels (red solid lines), without sub-Ch-mass SNe Ia (green short-dashed lines), and with 10 times more SNe Iax (blue long-dashed lines).
}
\end{figure}

\begin{figure}[t]
\center 
\includegraphics[width=8.5cm]{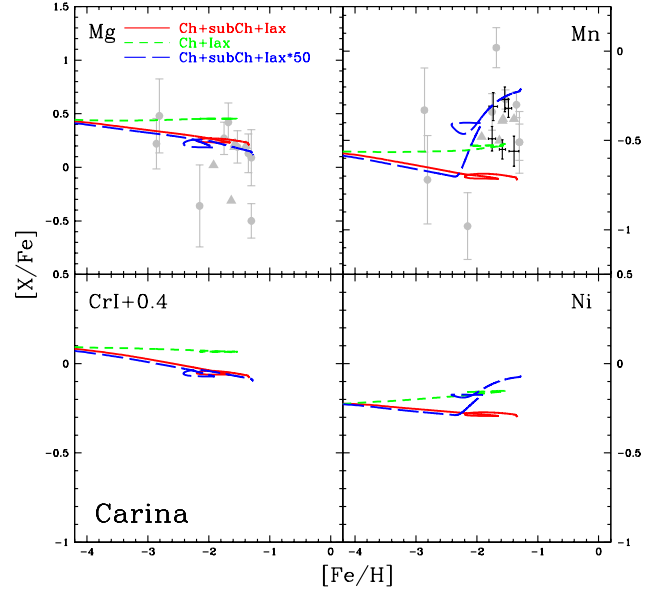}
\caption{\label{fig:carina}
The same as Fig.\ref{fig:fornax} but for Carina dSph galaxy with all SN Ia channels (red solid lines), without sub-Ch-mass SNe Ia (green short-dashed lines), and with 50 times more SNe Iax (blue long-dashed lines).
}
\end{figure}

Figures \ref{fig:fornax}-\ref{fig:carina} show the evolutions of elemental abundances for these four dSph models, with varying the contributions of various types of SNe Ia.
As for the solar neighborhood models, the elemental abundance ratios at [Fe/H] $\ltsim -3$ are determined from the IMF-weighted yields of core-collapse supernovae. The predicted ratios cannot be rejected by these small sample of observations, but lower [Mg/Fe] ratios might be preferred, which could be produced with an IMF truncated at $\sim 20M_\odot$ \citep{kob06,Nomoto2013}.
Around [Fe/H] $\sim -3$, [X/Fe] starts to decrease in the models with sub-Ch-mass SNe Ia, while [X/Fe] stays constant without sub-Ch-mass SNe Ia.
This transition is caused by the shortest lifetime, which is set at 0.04 Gyr as in the ``observed'' delay-time distribution \citep{mao14} for sub-Ch-mass SNe Ia in our models (\S \ref{sec:snia}).
Around [Fe/H] $\sim -1$, [(Mn, Ni)/Fe] rapidly increases in the models with Ch-mass SNe Ia, while [X/Fe] stays constant without Ch-mass SNe Ia.
This transition is caused by the metallicity effect of Ch-mass SNe Ia in our models.

As in \citet{kob15}, we call pure deflagrations of hybrid C+O+Ne WDs ``SNe Iax'' and use the nucleosynthesis yields from \citet{fin14}, which can produce very high [Mn/Fe] ratios at [Fe/H] $\ltsim -1$.
In the fiducial models of this paper, the SN Iax rate is determined from the calculated mass range of hybrid WDs (K19), which is much narrower and gives a lower SN Iax rate than in \citet{kob15}. The normalization is given by the same binary parameters for Ch-mass C+O WDs; $b_{\rm RG}=0.02$ and $b_{\rm MS}=0.04$.

In Figure \ref{fig:fornax} for Fornax, the observed [Mg/Fe] ratios seem to decrease from [Fe/H] $\sim -2$ toward higher metallicities, which cannot be reproduced without sub-Ch-mass SNe Ia (blue long-dashed lines).
The model with only sub-Ch-mass SNe Ia (green short-dashed lines) cannot reproduce the monotonic increase of [Mn/Fe] from [Fe/H] $\sim -1$ to $\sim -0.5$, and thus it is necessary to include Ch-mass SNe Ia as well.
The model including all three SN Ia channels (red solid lines) are in good agreement not only with [Mg/Fe] but also with [Mn/Fe] ratios.
Some of the stars with very high [Mn/Fe] ratios at [Fe/H] $\ltsim -2$ might be locally enriched by SNe Iax.
The errors of the other iron-peak elements are too large to place further constraints.

Also in Figure \ref{fig:sculptor} for Sculptor, the observed [Mg/Fe] decreases at [Fe/H] $\gtsim -2$, which cannot be reproduced without sub-Ch-mass SNe Ia.
The observed [Mn/Fe] ratios are highest around [Fe/H] $\sim -1.5$, and then sharply decreases until [Fe/H] $\sim -1$. This is not reproduced with the model including all three SN Ia channels (red solid lines).
Obviously, without Ch-mass SNe Ia, [Mn/Fe] monotonically decreases (green dotted lines). The SN Iax is a potentially good source to reproduce this [Mn/Fe] evolution but a higher rate is required. If we multiply the SN Iax rate by a factor of 50, then, it is possible to reproduce the rapid increase of [Mn/Fe] from [Fe/H] $\sim -2$ to $\sim -1.5$ (blue short-dashed lines). 
Then, in order to reproduce the [Mn/Fe] decrease from [Fe/H] $\sim -1.5$ to $\sim -1$, it is better to exclude normal Ch-mass SNe Ia (magenta long-dashed lines).

In Figure \ref{fig:sextans} for Sextans, the observed [Mg/Fe] ratios cannot be reproduced without sub-Ch-mass SNe Ia (green short-dashed lines), but are in reasonably good agreement with the model including all three SN Ia channels (red solid lines). The observed [Mn/Fe] can be better explained if the SN Iax rate is boosted by 10 times (blue long-dashed lines).
In Figure \ref{fig:carina} for Carina, the observed scatters are larger than the ranges of [X/Fe] evolutions, and the inhomogeneous enrichment should be important \citep{ven12}.
Nonetheless, the model including all three SN Ia channels (red solid lines) is closer to the observed [Mg/Fe] ratios, while the model without sub-Ch-mass SNe Ia is better for the observed [Mn/Fe] ratios (green short-dashed lines).
Similar to Sculptor, with a 50$\times$ boosted SN Iax rate (blue long-dashed lines), the model is in good agreement with the observations both for [Mg/Fe] and [Mn/Fe].
In summary, for all four dSphs we have modelled, [Mg/Fe] ratios can be well reproduced with larger enrichment from sub-Ch-mass SNe Ia than in the solar neighborhood.
However, with sub-Ch-mass SNe Ia, [Mn/Fe] ratios become too low, which can be solved with additional enrichment from SNe Iax.

\begin{figure*}[t]
\center 
\includegraphics[width=15cm]{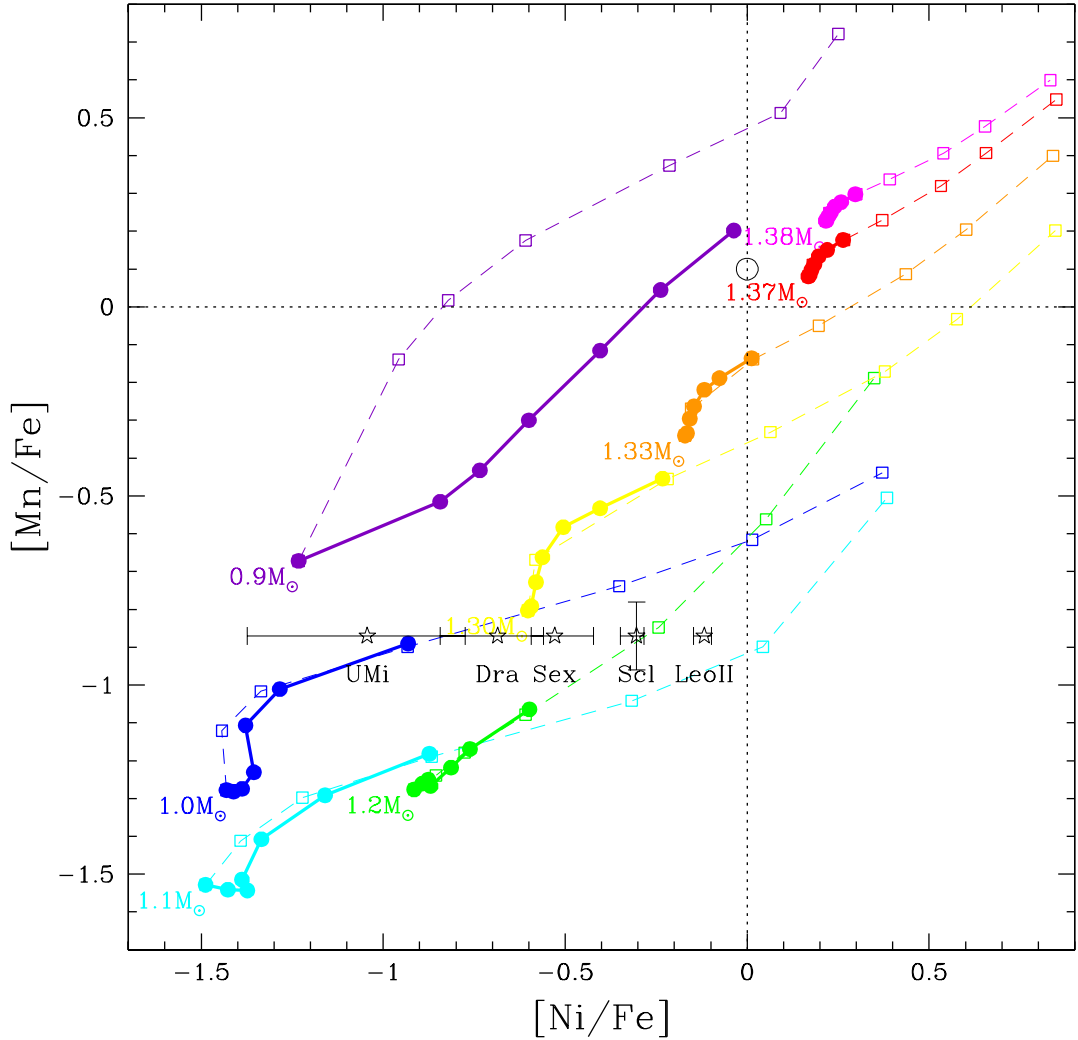}
\caption{\label{fig:nimn}
The Ni--Mn diagram for constraining the SN Ia enrichment.
Our nucleosynthesis yields are shown for
near-Ch-mass SNe Ia with WD masses $M_{\rm WD}=1.38$ (magenta), 1.37 (red), 1.33 (orange), and $1.30M_\odot$ (yellow) and initial metallicities $Z=0, 0.002, 0.01, 0.02, 0.04, 0.06$, and 0.10 (from left to right),
and for sub-Ch-mass SNe Ia with $M_{\rm WD}=0.9$ (purple), 1.0 (blue), 1.1 (cyan), and $1.2M_\odot$ (green) and $Z= 0, 0.001, 0.002, 0.004, 0.01, 0.02$, and 0.04 (from left to right).
Our yields with the solar-scaled initial composition (solid lines with filled circles) are significantly different from the LN18/LN19 yields with $^{22}$Ne only (dashed lines with open squares).
The large open circle indicates the average yields in the solar neighborhood at [Fe/H] $\gtsim -1$.
The stars with errorbars show the empirical yields obtained from the observed abundances of stars in dSphs from \citet{kir19}.
}
\end{figure*}

\subsection{The Mn/Fe--Ni/Fe diagram}

Mn and Ni are the key elements to constrain the enrichment sources in dSphs, and we present a useful diagram in Figure \ref{fig:nimn}.
We show our nucleosynthesis yields of near-Ch and sub-Ch mass models for various WD masses and initial metallicities in the diagram of [Mn/Fe] vs [Ni/Fe].
The near-Ch-mass models with different WD masses are calculated with changing central densities of WDs (LN18).
The solid lines are for the new yields in this paper, while the dashed lines are for LN18 and LN19 yields with only $^{22}$Ne for the $Z$ component of the initial composition.
There is an almost linear trend where both [Mn/Fe] and [Ni/Fe] increase with higher metallicities (see Fig. \ref{fig:comp_element_plot} for the reasons). At a given metallicity, [Mn/Fe] and [Ni/Fe] are higher for less massive WDs in sub-Ch-mass models, but for more massive WDs in near-Ch-mass models.

The dotted lines denote the solar ratios, and the large open circle indicates the average SN Ia yields in the solar neighborhood at [Fe/H] $\gtsim -1$.
It is clear that there is no single model that can simultaneously reproduce the Mn/Fe and Ni/Fe ratios, but the Ch-mass model (red solid line) is the closest.
$0.9 M_\odot$ sub-Ch-models (purple solid line) with the highest metallicity ($Z=0.04$) is also close, but $0.9 M_\odot$ WDs should be rare because of the low iron mass, and is even rarer at such high metallicities.

The stars with errorbars show the empirical SN Ia yields obtained from the observed evolutionary trends in dSphs.
Only Ni/Fe values are estimated for 5 dSphs (Scl, LeoII, Dra, Sex, UMi) \citep{kir19}, and the Mn/Fe value in Scl \citep{del19} is used for all dSphs in this figure.
The low Mn/Fe and Ni/Fe can be better explained with sub-Ch-mass models.
Note that, however, the initial composition of the nucleosynthesis calculation is crucial for this argument; with simplified models with only $^{22}$Ne, the observational data of dSphs could be well reproduced with low-metallicity sub-Ch-mass SNe Ia, while with more realistic solar-scaled initial composition, the dSphs data can be better reproduced with metal-rich sub-Ch-mass SNe Ia.
Normal Ch-mass SNe Ia ($1.37M_\odot$, red solid line) clearly cannot reproduce the dSphs data, which is consistent with our GCE results in Figs. \ref{fig:fornax}-\ref{fig:carina}.

\section{Conclusions}

In our quest to identify the progenitors of SNe Ia, we first update the nucleosynthesis yields both for Ch and sub-Ch mass C+O WDs, for a wide range of metallicity, with our two-dimensional hydrodynamical code \citep{Leung2015a} and the latest nuclear reaction rates.
In particular, new electron capture rates even change the W7 yields significantly for Cr, Mn, and Ni.
For the explosion mechanism, deflagration-detonation transition is used for Ch-mass SNe Ia (LN18), while the double detonation model with the carbon detonation triggered by helium detonation is used for sub-Ch-mass SNe Ia (LN19).
The helium envelope has to be as thin as $M({\rm He})=0.05M_\odot$; otherwise, Ti, V, and Cr would be over-produced at [Fe/H] $\gtsim -1.5$ (Fig. \ref{fig:final_sChand_std}).

We then include the nucleosynthesis yields in our galactic chemical evolution code \citep{kob00} to predict the evolution of elemental abundances in the solar neighborhood and dSph galaxies.
For Ch-mass SNe Ia, the timescale of supernovae is mainly determined from the metallicity-dependent secondary mass range of our single degenerate model (\citealt{kob98}; KN09).
For sub-Ch-mass SNe Ia, we use the delay-time distribution estimated from observed supernova rates \citep{mao14}.
Including failed supernovae, the star formation histories are assumed in order to reproduce other observational constraints such as the metallicity distribution functions (K19).

In the observations of the solar neighborhood stars, Mn shows an opposite trend to $\alpha$ elements, showing an increase toward higher metallicities, which is very well reproduced by Ch-mass SNe Ia, but never by sub-Ch-mass SNe Ia alone.
Mn is mainly produced by NSE during deflagrations in Ch-mass WDs where electron captures lower the electron fraction of the incinerated matter, and the double-detonation models for sub-Ch-mass WDs do not have enough material with such a low electron fraction.
A small amount of Mn can also be produced by incomplete Si-burning during detonations, depending on the initial metallicity.

Previously, the problem with Ch-mass SNe Ia was the over-production of Ni at high metallicities, which is not observed.
In this paper, however, we find that Ni yields of Ch-mass SNe Ia are much lower than in previous works when we use a more realistic initial composition of WDs (i.e., not $^{22}$Ne but the solar-scaled composition), which keeps the predicted Ni abundance within the observational scatter.
Among Ch-mass models, W7, 2D DDT, and 3D DDT give the elemental abundance ratios within the observational scatters in the solar neighborhood, and our 2D DDT gives the best fit to [Mn/Ni] ratios at $-1 \ltsim$ [Fe/H] $\ltsim 0.3$.
We also find that both for Ch and sub-Ch mass SNe Ia, the metallicity dependence of Mn and Ni is much weaker than in previous works (Fig. \ref{fig:comp_element_plot}).

From the evolutionary trends of elemental abundance ratios in the solar neighborhood, we conclude that the contribution of sub-Ch-mass SNe Ia in chemical enrichment is up to 25\%.
In dSph galaxies, however, the contribution of sub-Ch-mass SNe Ia seems to be higher than in the solar neighborhood, which is consistent with the low-metallicity inhibition of our single-degenerate scenario for Ch-mass SNe Ia.
In dSphs, sub-Ch-mass SNe Ia cause a decrease of [($\alpha$, Cr, Mn, Ni)/Fe], while so-called SNe Iax can increase Mn and Ni abundances {\it if} they are pure deflagrations.
Among dSphs, all galaxies we model in this paper (Fornax, Sculptor, Sextans, and Carina) seem to require larger enrichment from sub-Ch-mass SNe Ia than in the solar neighborhood. The observed [Mn/Fe] ratios in Sculptor and Carina may also require additional enrichment from SNe Iax.
Future observations of a large number of stars in dSphs would provide more stringent constraints on the progenitor systems and explosion mechanism of SNe Ia.

Within the one-zone GCE framework, it is not possible to reproduce the observed elemental abundance ratios of dSph stars (Figs. \ref{fig:fornax}-\ref{fig:carina}) only by the variations of IMF and SFRs among dSph galaxies.
Different SFRs could change the relative contribution between core-collapse supernovae and SNe Ia, and thus could change [Mg/Fe] ratios at a given time (or [Fe/H]). At the same time (or [Fe/H]), [(Cr, Mn, Ni)/Fe] ratios should also be similarly affected by SNe Ia. 
IMF variation mostly appears as the mass dependence of nucleosynthesis yields of core-collapse supernovae, and thus could change the normalization of abundance ratios. During core-collapse supernova explosions, Cr and Mn are synthesized in incomplete Si-burning regions, while Ni and Fe are synthesized in complete Si-burning regions, and thus [Cr/Mn] and [Ni/Fe] do not vary more than $\sim 0.2$ dex \citep{kob06}.
In this paper we show that the contributions from different sub-types of SNe Ia could explain the variations among [(Mg, Cr, Mn, Ni)/Fe] ratios.
Note that, however, in more realistic chemodynamical simulations, selective metal-loss could be caused by supernovae feedback in a shallow potential well, which might explain some of these variations in elemental abundances of dSph stars.

\acknowledgments
We thank K.\ Shen and I.\ Seitenzahl for providing nucleosynthesis data, and A.\ Ruiter for binary population synthesis data.
We are grateful to E. Kirby, M. de los Reyes, K. Hayashi, and A. Bunker for fruitful discussion.
CK acknowledges funding from the UK Science and Technology Facility Council (STFC) through grant ST/M000958/1 \& ST/R000905/1.
This work used the DiRAC Data Centric system at Durham University, operated by the Institute for Computational Cosmology on behalf of the STFC DiRAC HPC Facility (www.dirac.ac.uk). This equipment was funded by a BIS National E-infrastructure capital grant ST/K00042X/1, STFC capital grant ST/K00087X/1, DiRAC Operations grant ST/K003267/1 and Durham University. DiRAC is part of the National E-Infrastructure.
Numerical computations were also in part carried out on PC cluster at Center for Computational Astrophysics, National Astronomical Observatory of Japan.
SCL acknowledges support by funding HST-AR-15021.001-A.
This work has been supported by the World Premier International Research Center Initiative (WPI Initiative), MEXT, Japan, and JSPS KAKENHI Grant Numbers JP17K05382 and JP20K04024.

\appendix

\section{GCE Parameter Dependence}

\begin{figure*}
\center 
\includegraphics[width=15cm]{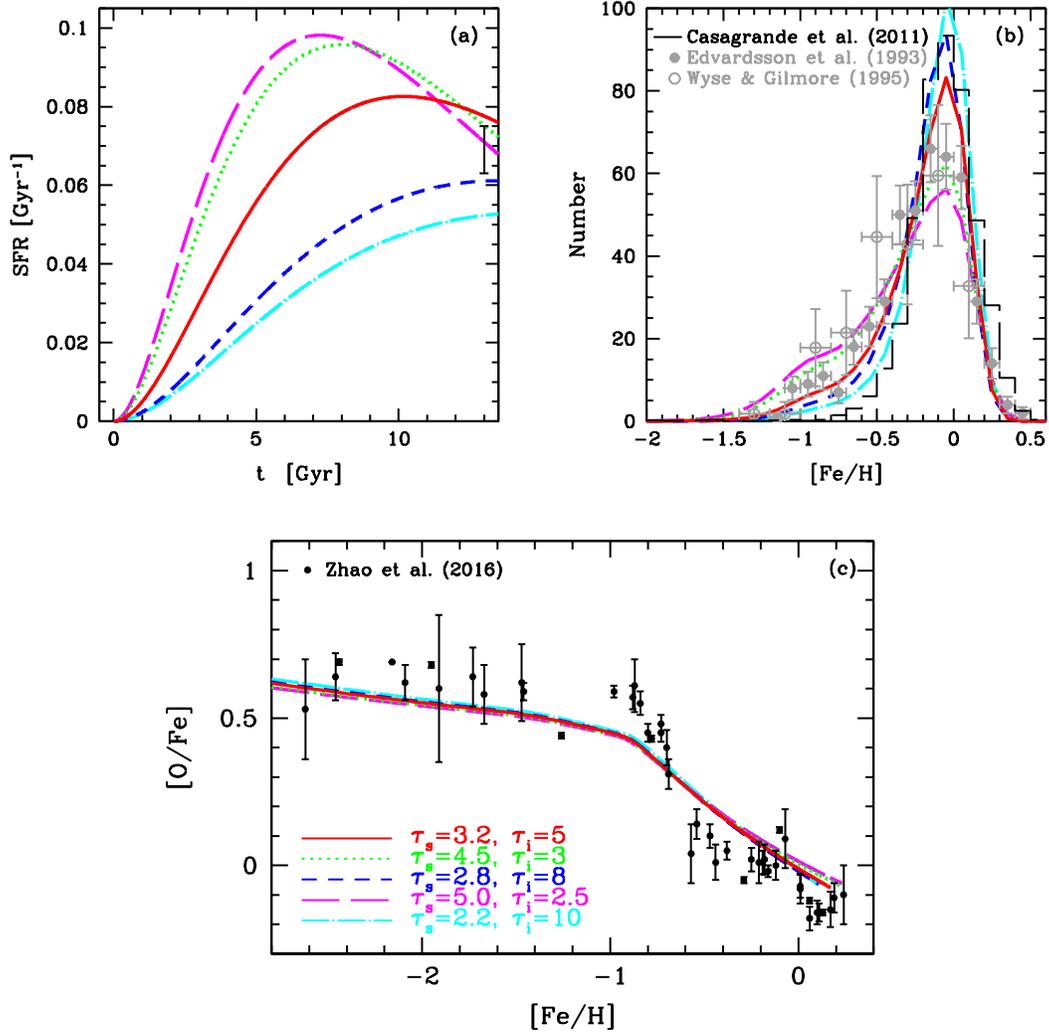}
\caption{\label{fig:gce2}
(a) Star formation rates, (b) metallicity distribution functions, and (c) [O/Fe]--[Fe/H] relations with a different set of star formation and infall timescales.
50\% Ch-mass $+$ 50\% sub-Ch-mass SN Ia contribution to GCE is assumed.
}
\end{figure*}

\begin{figure*}
\center 
\includegraphics[width=15cm]{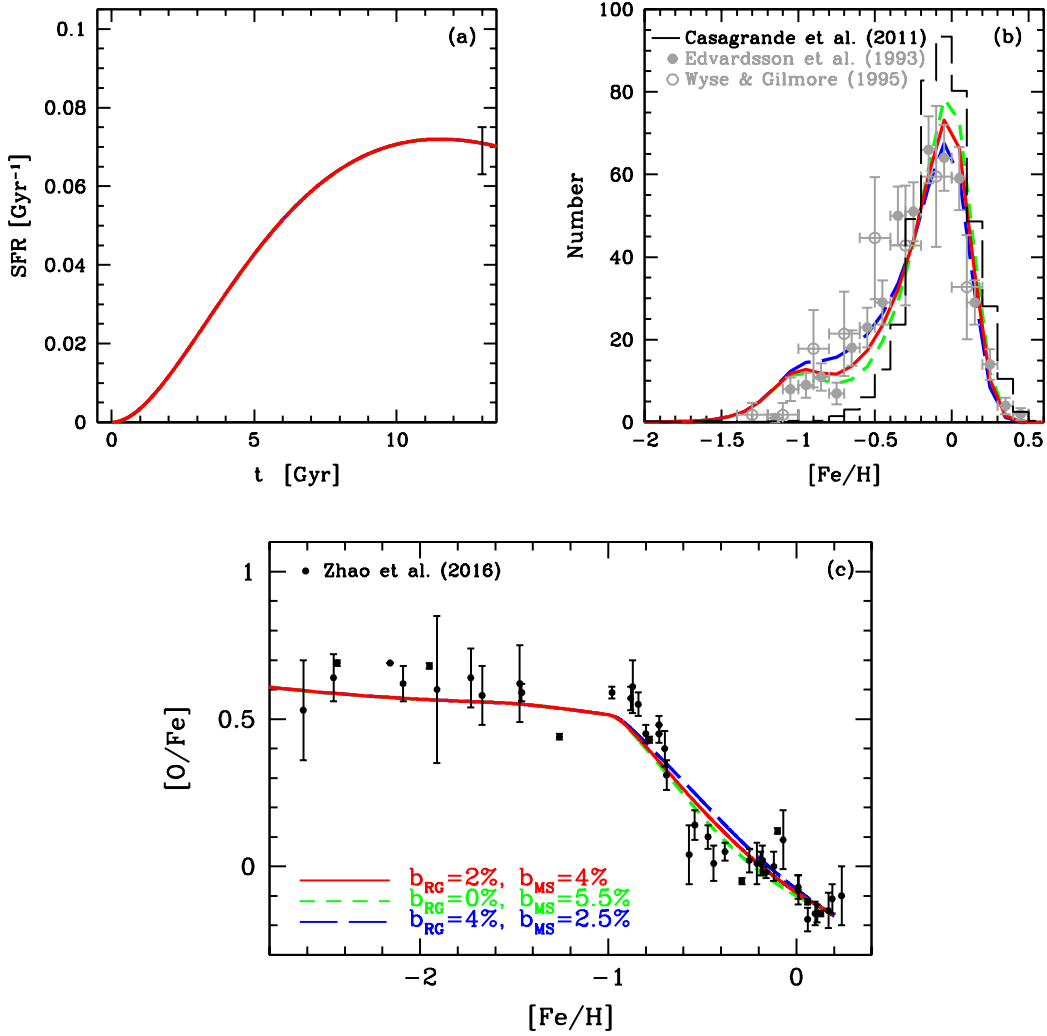}
\caption{\label{fig:gce3}
Same as Fig.\ref{fig:gce2} with a different set of binary parameters.
Only Ch-mass SN Ia is included.
}
\end{figure*}

In our solar neighborhood models, there are only two free parameters: the timescales of infall $\tau_{\rm i}$ and of star formation $\tau_{\rm s}$ in Gyrs (\S \ref{sec:gce}).
Figure \ref{fig:gce2} shows our chemical evolution models varying the two timescales, including 50\% sub-Ch-mass SNe Ia.
We choose the set of these timescales in order to match the observed MDF (panel b).
The peak [Fe/H] of the MDF provide `net' yields, which corresponds to the average stellar metallicity and depends only on nucleosynthesis yields, outflow (metal loss), and the IMF \citep{tin80}.
Therefore, $\tau_{\rm s}$ can be determined uniquely with a given $\tau_{\rm i}$. A shorter $\tau_{\rm s}$ value is required for a longer $\tau_{\rm i}$, which results in more rapid star formation (panel a).
However, the evolution of [O/Fe] as a function of [Fe/H] becomes almost identical, independent of the choice of these parameters.
This is because the MDF tells how quickly the star formation and chemical enrichment from core-collapse supernovae (which produce O) takes place, relative to the timescale of Ch and sub-Ch mass SNe Ia (which produce Fe).
This is why our conclusions using GCE models constrained with MDFs are robust.
In this paper, in order to compare the models varying sub-Ch-mass SNe Ia, we choose the models with $\tau_{\rm i}=5$ Gyr in Table \ref{tab:param}.

In our Ch-mass SN Ia model, the total number of SNe Ia are given by two binary parameters, $b_{\rm MS}$ and $b_{\rm RG}$ respectively for MS+WD and RG+WD systems (\S \ref{sec:snia}), and the ratio is theoretically uncertain.
Figure \ref{fig:gce3} shows our chemical evolution models varying the two binary parameters, including Ch-mass SNe Ia only.
We first choose the set of these binary parameters in order to reproduce the slope of [O/Fe] ratios against [Fe/H] (panel c). 
A smaller $b_{\rm MS}$ value (blue long-dashed lines) gives a slightly shallower curve of the [O/Fe]--[Fe/H] relation.
Note that since the time-delay is different for MS+WD and RG+WD systems, the total number of SNe Ia exploded by present is not simply the summation of the two numbers.
We then choose the fiducial values from the shape of the MDF; a smaller $b_{\rm MS}$ value gives a larger number of metal-poor stars because the iron production becomes slower on the average.
The model with $b_{\rm RG}=2\%$ and $b_{\rm MS}=4\%$ gives the best match with the observed MDF in the solar neighborhood, and thus we use this set for the fiducial model.

For dSph galaxies, the peak [Fe/H] of the MDFs is lower than in the solar neighborhood, which requires outflows, provided the same nucleosynthesis yields and IMF.
We consider two forms of outflows: one is the outflow proportional to the SFR, described by a timescale $\tau_{\rm o}$, and the other is the galactic wind set by $t_{\rm w}$ (\S \ref{sec:dsph}).
The value of $\tau_{\rm o}$ is chosen to match the peak [Fe/H] of the MDF (Fig. \ref{fig:dsph-mdf}c), while the value of $t_{\rm w}$ is chosen to reproduce the lack of young stars in the observations (Fig. \ref{fig:dsph-mdf}a).
Although there are more free parameters, because there are more observational constraints for dSphs than in the solar neighborhood, it is possible to choose the best parameter set.
In dSphs, SFRs are estimated not only at present but for the entire history, and from the shape $\tau_{\rm i}$ and $\tau_{\rm s}$ are constrained.
In general, slower star formation (with longer $\tau_{\rm i}$ and/or $\tau_{\rm s}$) leads to too low SFRs at $t \ltsim 4$ Gyr, while faster star formation (with shorter $\tau_{\rm i}$ and/or $\tau_{\rm s}$) leads to too low SFRs at $t \gtsim 4$ Gyr.

\bibliography{ms}{}
\bibliographystyle{apj}

\end{document}